\documentclass[fleqn,usenatbib]{mnras}
\usepackage{newtxtext,newtxmath}
\usepackage{booktabs}
\usepackage{longtable}
\usepackage[dvipsnames]{xcolor}
 \usepackage[T1]{fontenc}

\usepackage[T1]{fontenc}

\DeclareRobustCommand{\VAN}[3]{#2}
\let\VANthebibliography\thebibliography
\def\thebibliography{\DeclareRobustCommand{\VAN}[3]{##3}\VANthebibliography}

\usepackage{graphicx}	% Including figure files
\usepackage{amsmath}	% Advanced maths commands
\usepackage{ulem}
\usepackage{booktabs}
\newcommand{\nKnownQSOHighzWithGaiaSpec}{938} % Includes ALL QSOs in the QDB (T1, T2, BLLac) with z > 2.5 with a low res spectrum
\newcommand{\nRemainingCandidates}{5,469\ }
\newcommand{\nCandidatesWithGaiaSpec}{2,635\ }
\newcommand{\nNewRedshifts}{{1672} }
\newcommand{\nGaiaSpecNoz}{963\ } % 2635-1672
\newcommand{\nObsWithGaiaSpec}{{25\ }}
\newcommand{\nCandidatesLeft}{{3797\ }}
\newcommand{\nAllObjWithzSpec}{821,992} % Objects that satisfy: FROM Qubrics.All_info WHERE z_spec IS NOT NULL
\newcommand{\sv}{\sigma_v}
\def\cms{$ \rm cm~s^{-1}$}

\newcommand{\nqso}{N_{\rm QSO}}
\newcommand{\ti}{t_{\rm int}}
\newcommand{\zqso}{z_{\rm QSO}}
\newcommand{\zsp}{z_{\rm spec}}
\newcommand{\zQUG}{z_{\rm QU\_G}}
\def\mincir{\lower.5ex\hbox{$\; \buildrel < \over \sim \;$}}
\def\magcir{\lower.5ex\hbox{$\; \buildrel > \over \sim \;$}}
\title[Gaia Spectroscopy of QUBRICS QSO candidates]{Spectroscopy of QUBRICS quasar candidates: \nNewRedshifts new redshifts and a Golden Sample for the Sandage Test of the Redshift Drift}

\author[Cristiani et al.]{
\parbox[t]{\textwidth}{
Stefano Cristiani$^{1,2,3}$ \thanks{E-mail: stefano.cristiani@inaf.it},
Matteo Porru$^{1}$,
Francesco Guarneri$^{1,4,5}$,
Giorgio Calderone$^{1}$,
Konstantina Boutsia$^{6}$,
Andrea Grazian$^{7}$,
Guido Cupani$^{1}$, 
Valentina D'Odorico$^{1,2,8}$,
Fabio Fontanot$^{1,2}$,
Carlos J.A.P. Martins$^{9,10}$,
Catarina M. J. Marques$^{9,10,11}$,
Soumak Maitra$^{1}$,
Andrea Trost$^{1,4}$}
\vspace*{6pt}\\
$^{1}$ INAF--Osservatorio Astronomico di Trieste, Via G.B. Tiepolo, 11, I-34143 Trieste, Italy \\
$^2$ IFPU--Institute for Fundamental Physics of the Universe, via Beirut 2, I-34151 Trieste, Italy \\
$^3$ INFN--National Institute for Nuclear Physics,  
via Valerio 2, I-34127 Trieste, Italy \\
$^{4}$ Dipartimento di Fisica, Sezione di Astronomia, Universit\`{a} di Trieste, via G.B. Tiepolo 11, I-34131, Trieste, Italy \\
$^5$ ESO--European Southern Observatory, Karl-Schwarzschild-Strasse 2, 85748 Garching bei M\"{u}nchen, Germany \\
$^6$ Las Campanas Observatory, Carnegie Observatories, 
Colina El Pino, Casilla 601, La Serena, Chile\\
$^7$ INAF--Osservatorio Astronomico di Padova,
Vicolo dell'Osservatorio 5, I-35122 Padova, Italy \\
$^8$ Scuola Normale Superiore, P.zza dei Cavalieri, I-56126 Pisa, Italy\\
$^9$ Centro de Astrof\'{\i}sica da Universidade do Porto, Rua das Estrelas, 4150-762 Porto, Portugal \\
$^{10}$ Instituto de Astrof\'{\i}sica e Ci\^encias do Espa\c{c}o, CAUP, Rua das Estrelas, 4150-762 Porto, Portugal \\
$^{11}$ Faculdade de Ci\^encias, Universidade do Porto, Rua do Campo Alegre, 4150-007 Porto, Portugal
}

\date{Accepted 2023 March 30. Received 2023 March 10; in original form 2023 January 27.}

\pubyear{2023}

\begin{document}
\label{firstpage}
\pagerange{\pageref{firstpage}--\pageref{lastpage}}
\maketitle

\begin{abstract}
The QUBRICS (QUasars as BRIght beacons for Cosmology in the Southern hemisphere) survey aims at constructing a sample of the brightest quasars with $z \magcir 2.5$, observable with facilities in the Southern Hemisphere.
QUBRICS makes use of the available optical and IR wide-field surveys in the South and of Machine Learning techniques to produce thousands of bright quasar candidates of which
only a few hundred have been confirmed with follow-up spectroscopy.
Taking advantage of the recent Gaia Data Release 3, which contains 220 million low-resolution spectra, and of a newly developed spectral energy distribution fitting technique, designed to combine the photometric information with the Gaia spectroscopy, it has been possible to measure \nNewRedshifts new secure redshifts of QUBRICS candidates, with a typical uncertainty of $\sigma_z = 0.02$.
This significant progress of QUBRICS brings it closer to (one of) its primary goals: providing a sample of bright quasars at redshift $2.5<z<5$ to perform the Sandage test of the cosmological redshift drift.
A Golden Sample of seven quasars is presented that makes it possible to carry out this experiment in about 1500 hours of observation in 25 years, using the ANDES spectrograph at the 39m ELT, a significant improvement with respect to previous estimates.
\end{abstract}


%%%%%%%%%%%%%%%%%%%%%%%%%%%%%%%%%%%%%%%%%%%%%%%%%%

%%%%%%%%%%%%%%%%% BODY OF PAPER %%%%%%%%%%%%%%%%%%

\section{Introduction}
Quasars(QSOs), as the brightest non-transient sources, can be observed at very high redshifts and shed light on fundamental topics such as the formation and evolution of galactic structures and massive black holes, the Big Bang nucleosynthesis, Cosmology, reionizations, and the variation of the fundamental constants. 
As cosmic lighthouses, they provide a unique view of the Universe through the observation of absorption features, 
and the brightest QSOs are coveted as precious tools of investigation.

The QUBRICS (QUasars as BRIght beacons for Cosmology in the Southern hemisphere) survey \citep{Calderone19:2019ApJ...887..268C}
has been conceived with the aim of making up for the scarcity of bright QSOs in the Southern Hemisphere,
which is due to the historical paucity of all-sky surveys in the South, and
has produced several hundreds new spectroscopically confirmed bright QSOs.
Various methods for the selection of QSOs have been used:
in \citet[hereafter Paper I]{Calderone19:2019ApJ...887..268C} candidates have been selected using a canonical correlation analysis \citep[CCA, ][]{ref:CCA},
in \citet[hereafter Paper III]{Guarneri:2021MNRAS.506.2471G} the Probabilistic Random Forest \citep[PRF, ][]{ReisPRF:2019AJ....157...16R} has been adopted, with modifications introduced to properly treat upper limits and missing data.
In \citet[Paper VI]{Guarneri2022} the PRF selection has been further improved, in particular adding synthetic data to the training sets. 
Calderone et al., (in preparation, Paper VII) have developed a method that takes advantage of the extreme gradient boosting technique \citep[XGB]{XGBoost2016} to significantly improve the recall
\footnote{{\it Recall} is defined as the fraction of relevant instances that are retrieved by the selection algorthm.} of the  selection algorithms even in the presence of severely imbalanced datasets, with the aim of completing the QUBRICS survey up to $z \sim 5$.

While refining the methods of selection, a continuous effort has been dedicated in QUBRICS to the follow-up spectroscopy \citep[hereafter Paper II]{Boutsia2020}, testing the selection procedures and
leading to statistically well-defined subsamples that allowed us to address the topics of the QSO luminosity function (LF) and cosmic re-ionization(s)
(\cite{LF_Boutsia:2021ApJ...912..111B}, hereafter Paper IV; \cite{LF_Hz_Grazian:2021arXiv211013736G}, Paper V; and \cite{Fontanot2023}, Paper IX).
Rare objects, such as extreme broad absorption line QSOs (BALs), discovered in the course of QUBRICS, have been described in \citet[Paper VIII]{ref:cupaniFeLoBALs}.
One of the main goals of QUBRICS is to provide a sample of bright targets for the Sandage test of the cosmological redshift drift \citep{Sandage62}.
The redshift drift ($\dot{z} = {{\rm d}z}/{{\rm d}t_{\rm obs}}$) is a small, dynamical change in the redshift of objects following the Hubble flow. Measuring it provides a direct, real-time and model-independent mapping of the expansion rate of the universe. It is fundamentally different from other cosmological probes: instead of mapping our (present-day) past light cone, it directly compares different past light cones. Being independent of any assumptions on gravity, geometry or clustering, it directly tests the pillars of the $\Lambda$CDM paradigm. Recent theoretical studies have uncovered unique synergies with other cosmological probes, in particular for the characterization of the physical properties of dark energy \citep{Martins2016, Alves2019, Esteves2021}. This measurement is a flagship objective of the Extremely Large Telescope (ELT) \citep{Liske+08:2008MNRAS.386.1192L}, specifically observing the Lyman forest of QSOs with its high-resolution spectrograph, ANDES \citep{ANDES_MArconi2022}. The effect is tiny,
expected to be of order \cms yr$^{-1}$ at
the redshifts of interest, and to carry out the measurement a high signal to noise ratio (SNR) is a necessary condition, implying, even with bright QSOs and large telescopes, a huge investment of observing time.
Having the brightest possible cosmic beacons is a key ingredient to make it possible to perform a measurement which is presently at the edge of feasibility.
Since the observability of the Lyman forest with ground-based spectrographs requires a substantial redshift, the QUBRICS has been focused on QSOs at  $z \geq 2.5$.
\begin{figure*}
    \centering
    \includegraphics[width=\textwidth]{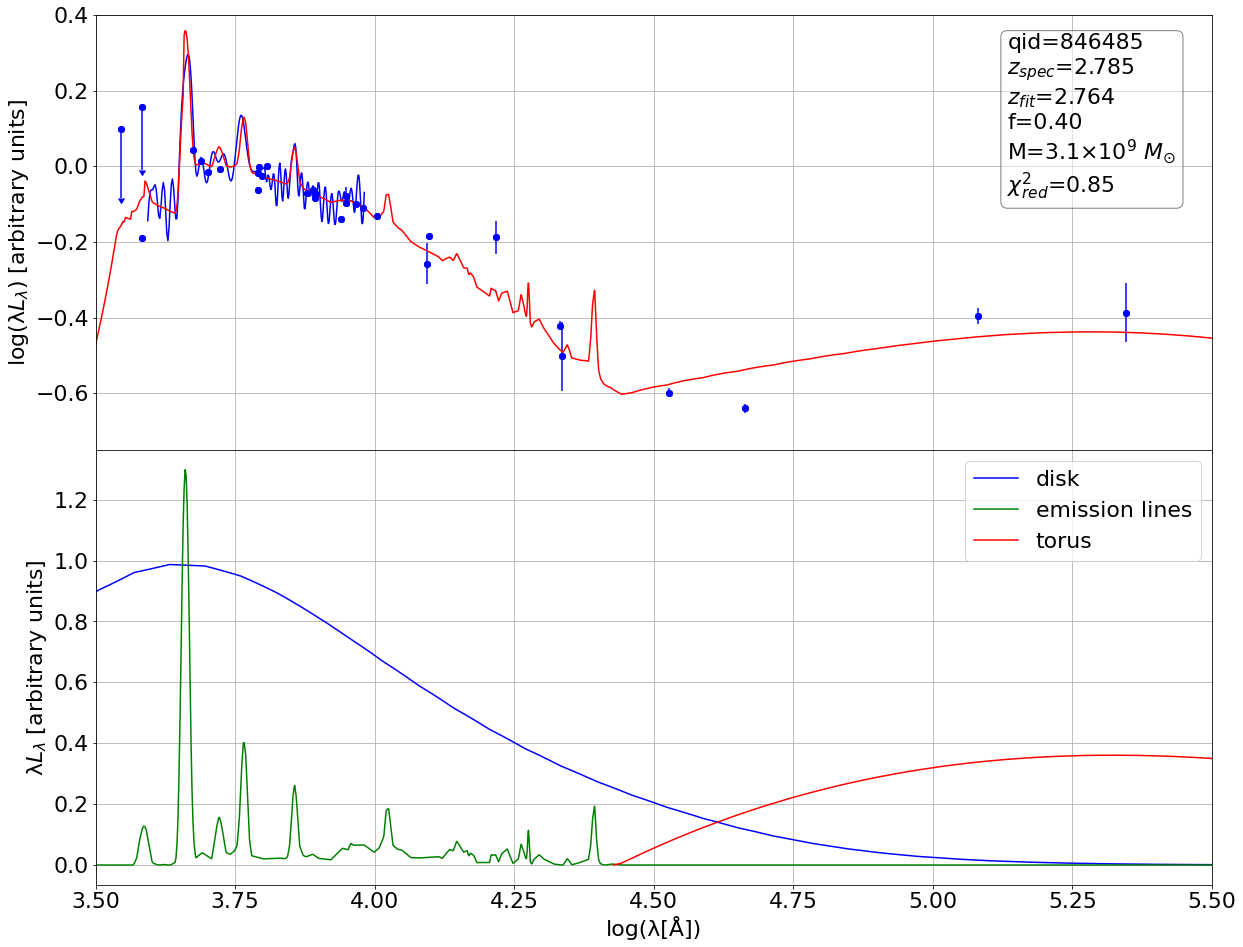}
    \caption{An example of SED fit (red line, upper panel) for a QSO with spectroscopic redshift $z_{\rm spec}=2.785$. In the upper panel the blue points show the photometric data, while the blue line is the Gaia low-resolution spectrum.
    The parameterization of the SED is described in the text and the resulting optimal parameters, $f$ (adimensional) and $M$ (in solar masses), are shown in upper right part of the figure. "qid", here and in the following, is the unique identifier of each source in the QUBRICS database.
    \label{Fig:SED_fit}
}
\end{figure*}

A strategic feature of QUBRICS is the continuous updating, after each observation cycle, of the training set, paying attention to identify and correct the non-insignificant fraction of erroneous spectroscopic identifications found in the literature, which may affect the training of machine learning (ML) techniques. This allows us to improve the success rate and the completeness, while keeping the list of candidates manageable.

In June 2022, the list of candidates derived from Papers III, VI and VII still lacking a spectroscopic confirmation included \nRemainingCandidates{} targets.
On June 13, 2022 the Gaia Data Release 3 (DR3) was published, providing, among a wealth of data, low-resolution spectra for about 220 million objects, selected to have a reasonable number of Gaia observations and to be sufficiently bright to ensure good SNR
\citep{GaiaSp2022}.
In this paper we make use of the DR3 spectra and of the QUBRICS photometric database (Sect.~\ref{sec:SED_fitting}),
combined with a spectral energy distribution (SED) fitting technique,
to obtain secure spectroscopic identifications and redshifts
of a significant fraction of the \nRemainingCandidates{} 
candidates of the QUBRICS survey.

Unless stated otherwise, magnitudes are given in the AB magnitude system; uncertainties represent 68\% confidence intervals. We adopt a flat $\Lambda$CDM cosmology, with $\Omega_\mathrm{m}=0.31,\ \Omega_\Lambda = 0.69$, and $H_0=67.7\ \mathrm{kms^{-1}Mpc^{-1}}$, in agreement with the \cite{Planck18}.
\section{Matching the QUBRICS database to the Gaia DR3 low-resolution spectra.}
\label{sec:QubricsDB}
The QUBRICS database is made of a collection of spectroscopic and photometric data from the literature, and all spectroscopic follow-ups carried out in the framework of the QUBRICS survey. The photometric database includes optical and infrared data from several public catalogs:
\begin{enumerate}
    \item $u$, $v$, $g$, $r$, $i$, $z$ magnitudes from the SkyMapper DR1 survey \citep{SkyMapper1:2018PASA...35...10W}%, SkyMapper3:2019PASA...36...33O};
    \item $G$, $G_{BP}$, $G_{RP}$ magnitudes from the Gaia eDR3 catalogue \citep{GaiaEDr3:2021A&A...649A...1G};
    \item $J$, $H$, $K$ from 2MASS \citep{2MASS:2006AJ....131.1163S};
    \item W1, W2, W3, W4 from the AllWise survey \citep{WISE:2010AJ....140.1868W};
    \item $g$, $r$, $i$, $z$, $Y$ magnitudes from the PanSTARRS1 DR2 survey \citep{PanSTARRS:Chambers_2016};
    \item $g$, $r$, $i$, $z$, $Y$ magnitudes from the DES survey \citep{DESY3Gold};
\end{enumerate}
From these datasets we extract convenient subsets on which the selection is performed, and we refer to them as Main Samples (MS). MS are created based on magnitude and data-quality cuts, and positional match distance. For instance, the MS used in \cite{Calderone19:2019ApJ...887..268C} includes sources with i) a magnitude measurement in six bands (SkyMapper \textit{i, z}, WISE W1, W2, W3 and Gaia G); ii) a magnitude $i$ limited between $14<i<18$; iii) ${\rm SNR} > 3$ in the AllWise bands; iv) matching distance between Gaia and SkyMapper, and AllWise and SkyMapper lower than 0.5''.

The Gaia catalogue additionally provides parallax and proper-motion measurements. This information is used to identify stars, which are assumed to be objects with parallax and proper motion significantly different from zero ($>3\sigma$).

Spectroscopic redshifts and classifications are collected from several catalogues and added to the database: 
\begin{itemize}
    \item the Sloan Digital Sky Survey (SDSS) DR16q \citep[][]{LykeSDSS16q:2020ApJS..250....8L};
    \item the Veron-Cétty catalogue \citep{Veron10:2010A&A...518A..10V}, including some of the correction by \citet{FleschVeronCorrections};
    \item the 2dF \citep{2df:2001MNRAS.328.1039C};
    \item the 6dF \citep{6df:2009MNRAS.399..683J};
    \item \citet{Onken21:2021arXiv210512215O};
    \item \citet{SDSSIncomplete_Schindler_PS:2019ApJ...871..258S};
    \item \citet{SDSSIncomplete_Schindler:2019ApJ...871..258S};
    \item \citet{Wolf20QSO:2020MNRAS.491.1970W};
    \item \citet{paper:yangQSO}.
\end{itemize}
Finally, 
QUBRICS identifications are added to the database after every observing run.
Combining the contribution of all these catalogues provides us, as of the end of June 2022, \nAllObjWithzSpec{} objects with a secure redshift estimate and classification (Tab. \ref{tab:objectSubdivision}).

These identifications are used to train the selection algorithms (CCA, PRF, XGB), which are then applied on unclassified sources to 
find QSO candidates. Combining the lists produced by the three algorithms leaves \nRemainingCandidates{} candidates still lacking a spectroscopic confirmation.
These candidates are scattered over a large area of the sky and observing all of them is a daunting task.

The publication of Gaia low-resolution spectra gave us the chance to
significantly speed up this process.
We cross-matched the candidates list to the Gaia DR3 source table, using a 0.75'' matching radius:
\nCandidatesWithGaiaSpec{} of the \nRemainingCandidates{} candidates turned out to have a low-resolution spectrum available from the Gaia archive. 

The Gaia spectra have been processed using Gaiaxpy\footnote{\url{https://gaia-dpci.github.io/GaiaXPy-website/}, 10.5281/zenodo.6637762}: each spectrum was calibrated and sampled on a wavelength grid using the \texttt{calibrate} routine; we chose a fixed step of 20 \AA{} and discarded the regions with wavelength < 3900 and > 9600 \AA{} , as these are typically very noisy.

The goal of the next Section is to use these data to obtain a secure redshift for a large fraction of the \nCandidatesWithGaiaSpec{} QUBRICS candidates
with a Gaia spectrum.

\begin{table}
    \centering
    \begin{tabular}{l|r|c}
    \toprule
        Object & Number of & Observed by \\
        type  & redshifts & QUBRICS \\
        
    \midrule
        QSO (Type 1) & 781298 & 856 \\
        QSO (Type 2) & 6074   & --- \\
        BLLac        & 809    & --- \\
        Galaxy       & 33614  & 23  \\
        Star         & 31     & 31  \\
    \bottomrule
    \end{tabular}
    \caption{Breakdown of sources with secure spectroscopic redshifts in the QUBRICS database. Spectroscopically confirmed stars are few compared to QSOs or galaxies: stars are identified mostly using Gaia proper motion and parallax criteria, and are not followed up spectroscopically. The 31 observed stars are misclassified as QSOs by our selection algorithms.}
    \label{tab:objectSubdivision}
\end{table}
\begin{figure}
    \centering
    \includegraphics[width=9.05cm]{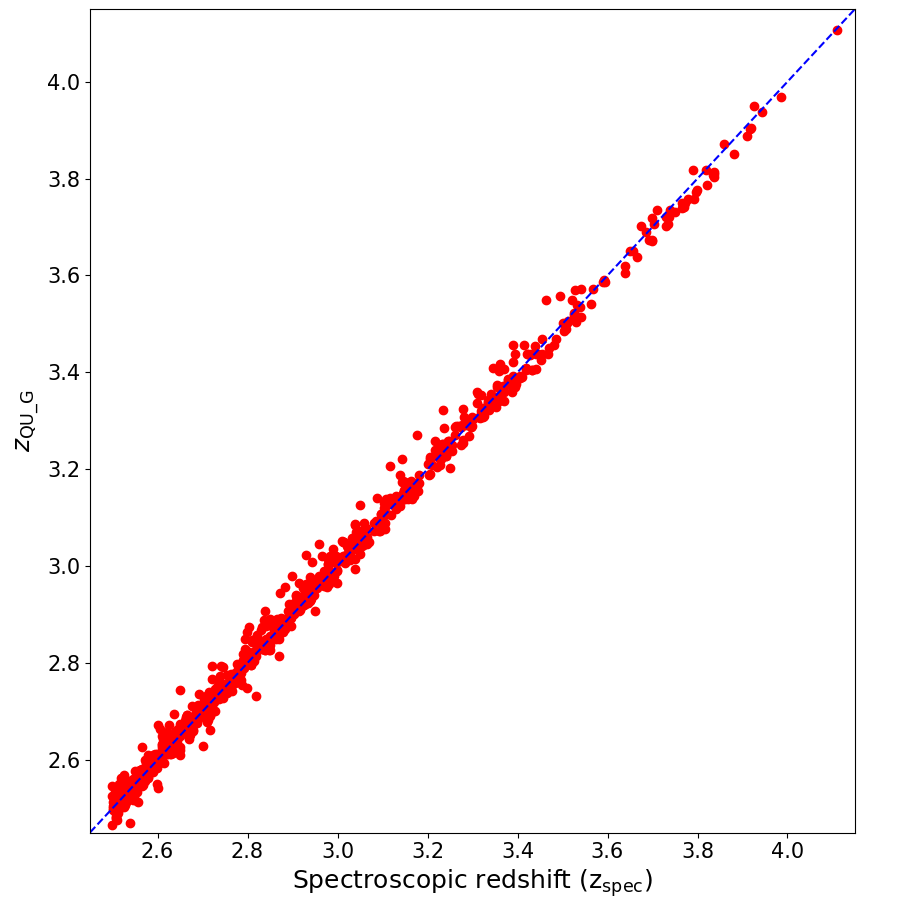}
    \includegraphics[width=8.8cm]{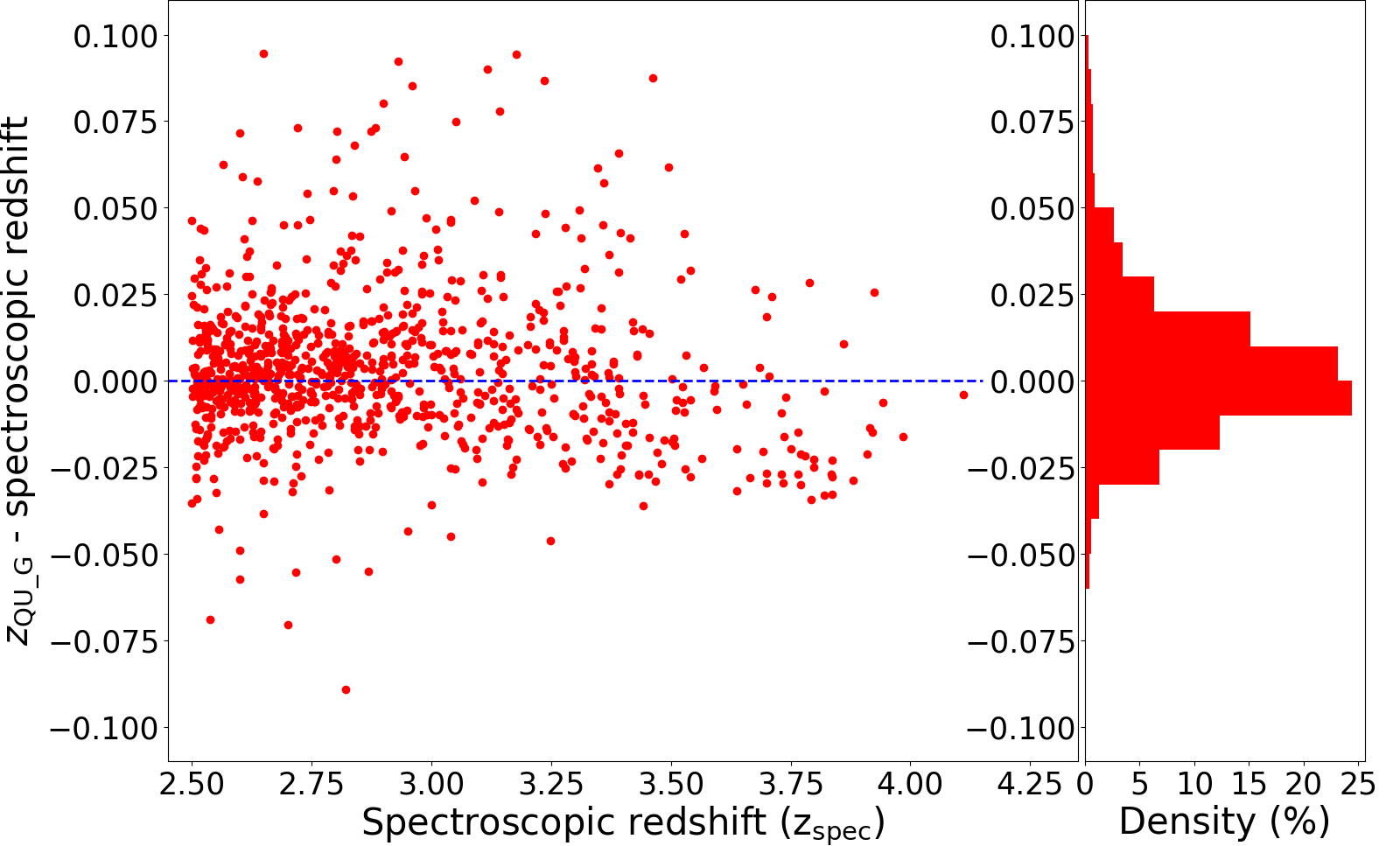}
    \caption{Top Panel: the redshifts determined on the basis the Gaia low-resolution spectra with the procedure described in Sect.\ref{sect:z_from_Gaia} vs. the spectroscopic redshifts for \nKnownQSOHighzWithGaiaSpec{} known QSOs with $z>2.5$ in the QUBRICS database. 
    Bottom panel: the difference between the redshifts determined on the basis the Gaia low-resolution spectra and the spectroscopic redshifts, $\Delta z$, as a function of the spectroscopic redshift, $\zsp$. 
    }
    \label{Fig:Dz_test_sample}
\end{figure}
\section{Obtaining secure spectroscopic identifications}
\label{sect:z_from_Gaia}
Having a set of \nCandidatesWithGaiaSpec{} objects confidently classified as $z \magcir 2.5$ QSOs \citep{Guarneri:2021MNRAS.506.2471G, Guarneri2022}, we derive, for each of them, the redshift $z$, comparing a robust model of the SED to their spectro + photometric data in the following way (described in detail in Sect.~\ref{sec:SED_fitting} and \ref{sec:MARZfit}.):
 \begin{enumerate}
    \item for each object an approximate redshift estimate is obtained by $\chi^2$-fitting a QSO SED
    to the photometric information available in QUBRICS complemented with the Gaia low-resolution spectroscopy;
    \item the redshift estimate is refined by processing the GAIA DR3 low-resolution spectrum with the MARZ package \citep{MARZ_2016}, which uses a cross-correlation algorithm to match the spectrum to a set of templates (in our case a QSO template).
\end{enumerate}
Consequently, three different types of redshift are used in this paper:
\begin{enumerate}
    \item  $\zsp$, the spectroscopic redshift obtained with follow-up spectroscopy, as, for example, described in Sect.~\ref{SEC:SpectrFollowUp}; 
    \item $\zQUG$, the redshifts derived from low-resolution Gaia spectra \citep{GaiaSp2022} combined with photometric data, as described in this section; 
    \item $z_{\rm Gaia}$, redshifts estimated by the Gaia Collaboration \citep{GaiaSp2022}, as reported in Sect.~\ref{sec:conclusions}.
\end{enumerate}
\subsection{SED fitting to estimate the redshift of a QSO}
\label{sec:SED_fitting}

For each object we compare the observed spectro+photometric data with synthetic SEDs of QSOs.

We use the same photometric catalogues listed in Sect. \ref{sec:QubricsDB}. In addition, when available, optical data from the SDSS DR16q survey \citep{LykeSDSS16q:2020ApJS..250....8L}, $J$, $H$, $K$ magnitudes from the VHS survey \citep{VHS:McMahon_2021} and $NUV$ magnitude from the GALEX survey \citep{GALEX:Morrissey_2007} are included. Finally, we add Gaia low-resolution spectra, rebinned with a step of 20 \AA~ in wavelength.

The SED of quasars is parametrized in the following way:
\begin{enumerate}
    \item a "blue bump" component, $F_{BB}(M, \dot{m}$), representing the accretion disk  \citep{Sun1989, Laor1989}, modeled with the PYAGN package \citep{PYAGN_2018}.
    The shape of this spectral component depends on the mass $M$ of the supermassive black hole (SMBH) and on its accretion rate $\dot{m}$: in particular, a larger $M$ tends to move the blue bump to longer wavelengths, while a higher $\dot{m}$ shifts it to shorter wavelengths;
    \item an "IR bump" component, $F_{IR}$, representing a dusty torus \citep{Pier1993, Mor2009}, starting from a rest wavelength of 8000  \AA~, that we
    derived from the QSO SED by \cite{Richards2006};
    \item a component representing the emission features, $F_{em}$, obtained with the QSFit package \citep{QSOFITS_2017} using the line fluxes and equivalent widths from \cite{VandenBerk2001} to normalize the intensity of each emission line with respect to the other lines and to the blue bump component. It should be noted that in this way the EW of the emission lines is fixed to the 
 \cite{VandenBerk2001} template. For QSOs with more exotic spectra (e.g. BALs) the procedure might result less reliable. This is further discussed in Sect.~\ref{sec:MARZfit}.
\end{enumerate}
Different SEDs can be generated by choosing a different SMBH mass ($M$), accretion rate ($\dot{m}$) and fractional contribution ($f$) of the "IR bump" with respect to the "blue bump" and emission features components,
according to the following formula:
\begin{equation}
   SED = (1-f) ( F_{BB}(M, \dot{m}) + F_{em} ) + f \cdot F_{IR}
   \label{eq:SED}
\end{equation}
Finally, according to the given redshift, the mean absorption of the intergalactic medium (IGM) is applied shortward of 1210 $\textup{\AA}$ \citep{Inoue2014}, and the SED is shifted to the observer's frame.

$F_{BB}$, $F_{em}$ and $F_{IR}$, dimensionally, are all fluxes. The resulting SED is renormalized and compared $\chi^2$-wise with the available spectro+photometric data. In this way a $\chi^2$ is obtained as a function of
the parameters $z$, $M$, $\dot{m}$ and $f$.
An example of a fit of a synthetic SED to the spectro+photometric data of a redshift $z_{\rm spec}=2.785$ QSO is shown in Fig.~\ref{Fig:SED_fit}.

\begin{figure}
    \centering
    \includegraphics[width=\hsize]{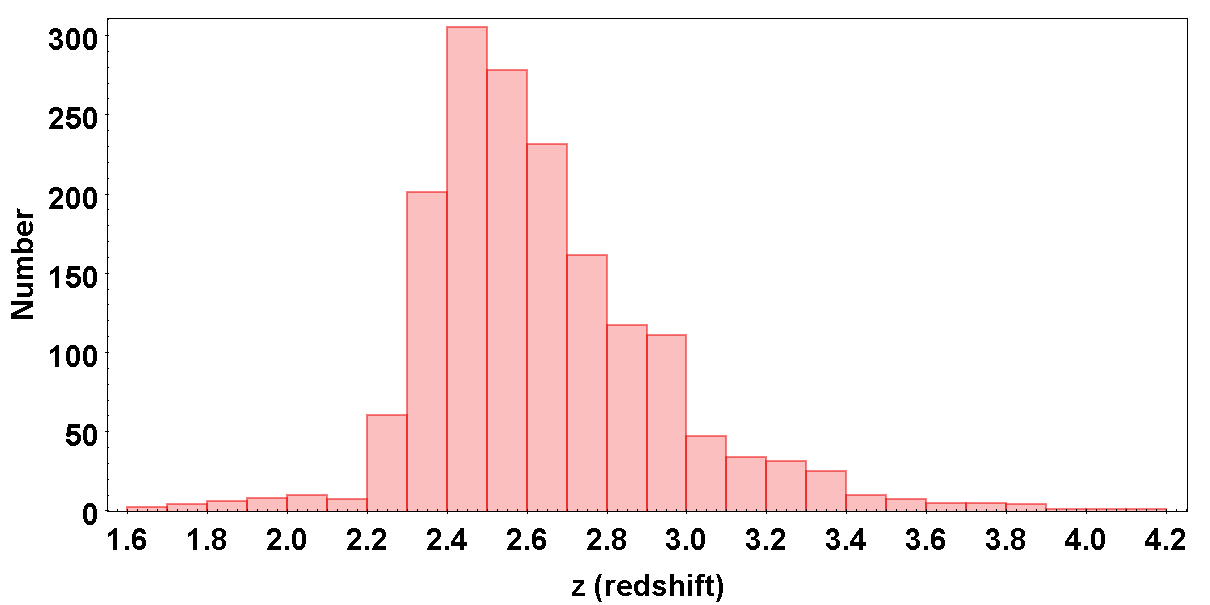}
    \caption{Histogram of the \nNewRedshifts new redshifts determined with Gaia spectroscopy
    }
    \label{Fig:z_hist_cand}
\end{figure}
\begin{figure}
    \centering
    \includegraphics[width=\columnwidth]{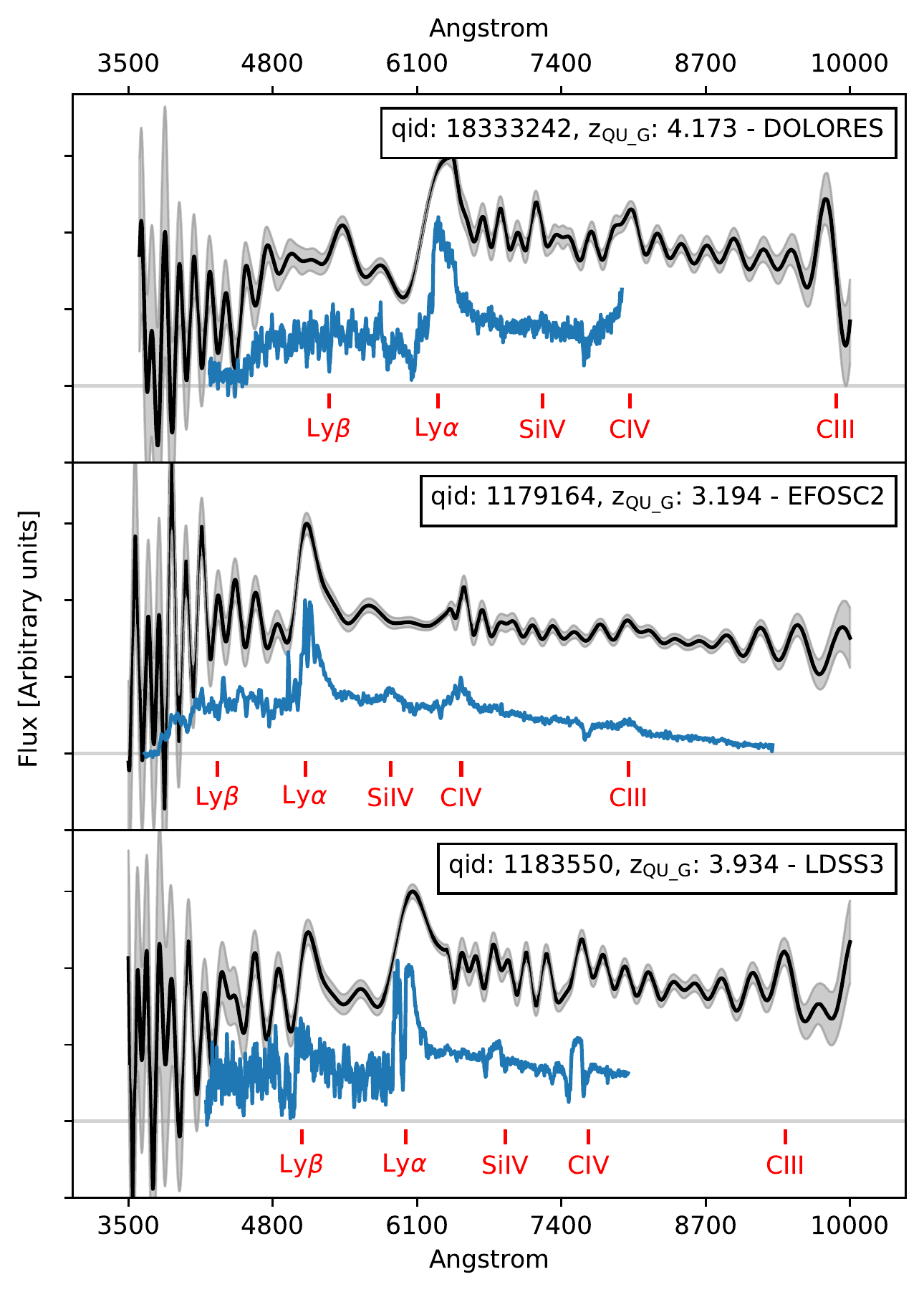}
    \caption{Example of discovery spectra for 3 of the QSO in the published catalogue. Spectra were taken with the DOLORES, EFOSC2 and LDSS3 spectrographs. In each panel, the black line shows the low-resolution Gaia spectrum, the blue line the spectrum taken at the telescope; prominent emission lines are marked in red. }
    \label{fig:exampleSpectra}
\end{figure}
To increase the robustness of the fitting procedure we explored whether the number of fitting parameters in Eq.~\ref{eq:SED} could be reduced.
Since the QUBRICS survey is focused on finding QSOs at $z>2.5$, we have extracted from the QUBRICS database the \nKnownQSOHighzWithGaiaSpec{}
spectroscopically confirmed QSOs  with $\zsp>2.5$ and a Gaia low-resolution spectrum. We fitted their spectro+photometric data  fixing the redshift at the spectroscopic value and letting the parameters $f$, $M$ and $\dot{m}$ vary.
Most QSO SEDs are fitted with an $\dot{m} \sim 1$ (and a non-negligible degeneracy exists between $M$ and $\dot{m}$). 
Therefore, we decided to fix $\dot{m}=1$ (i.e. Eddington accretion) in the following analysis and fit for each object three parameters: $f$, $M$ and the redshift $z_{\rm fit}$.

It should be noted that in the present context the aim of the spectro+photometric data-fitting is to provide a robust estimation of the redshift for sources classified as QSOs by our ML techniques. The focus is not on measuring other properties as the SMBH mass or accretion rate. The present parameterization of the QSO SED is also different with respect to \cite{Guarneri2022} and has been chosen because it produces better and more physically motivated results.

\subsection{Refining the redshift estimate by cross-correlation with the MARZ package}
\label{sec:MARZfit}
The redshift estimate derived from the SED fitting described in Sect.~\ref{sec:SED_fitting}
is finally passed to the MARZ package \citep{MARZ_2016}, that matches
the Gaia low-resolution spectrum of the object with the MARZ QSO template (see \citet{MARZ_2016} for the details of the MARZ matching technique) and produces our final estimate for the redshift, $\zQUG$. 
A quality operator (QOP, \cite{MARZ_2016}) is also assigned to each spectrum in a human-supervised way. The QOP scale varies from a value of 1 for inconclusive spectra to 4 for "great" spectra, with an absolutely certain redshift.

Fig.\ref{Fig:Dz_test_sample} shows the comparison between the spectroscopic redshifts, $\zsp$, present in the QUBRICS database and the $\zQUG$, as well as their difference, $\Delta z$, as a function of the spectroscopic redshift.
The distribution is characterized by a median $<\Delta z>= 0.001$
and a standard deviation $\sigma_{\Delta z} = 0.02$.
The combination of the photometric and spectrophotometric data and 
of the $\chi^2$ fitting and template matching techniques turns out to be effective in avoiding the misidentification of spectral features, while providing relatively precise redshifts.
In particular, potential errors due to QSOs with exotic spectra are checked and corrected/eliminated with the cross-correlation MARZ procedure and the reliability, tested on the sample of \nKnownQSOHighzWithGaiaSpec{} QSOs with known $\zsp$ (including a significant fraction of BALs), turns out to be reassuring (Fig.~\ref{Fig:Dz_test_sample}).
\begin{figure}
    \centering
    \includegraphics[width=\columnwidth]{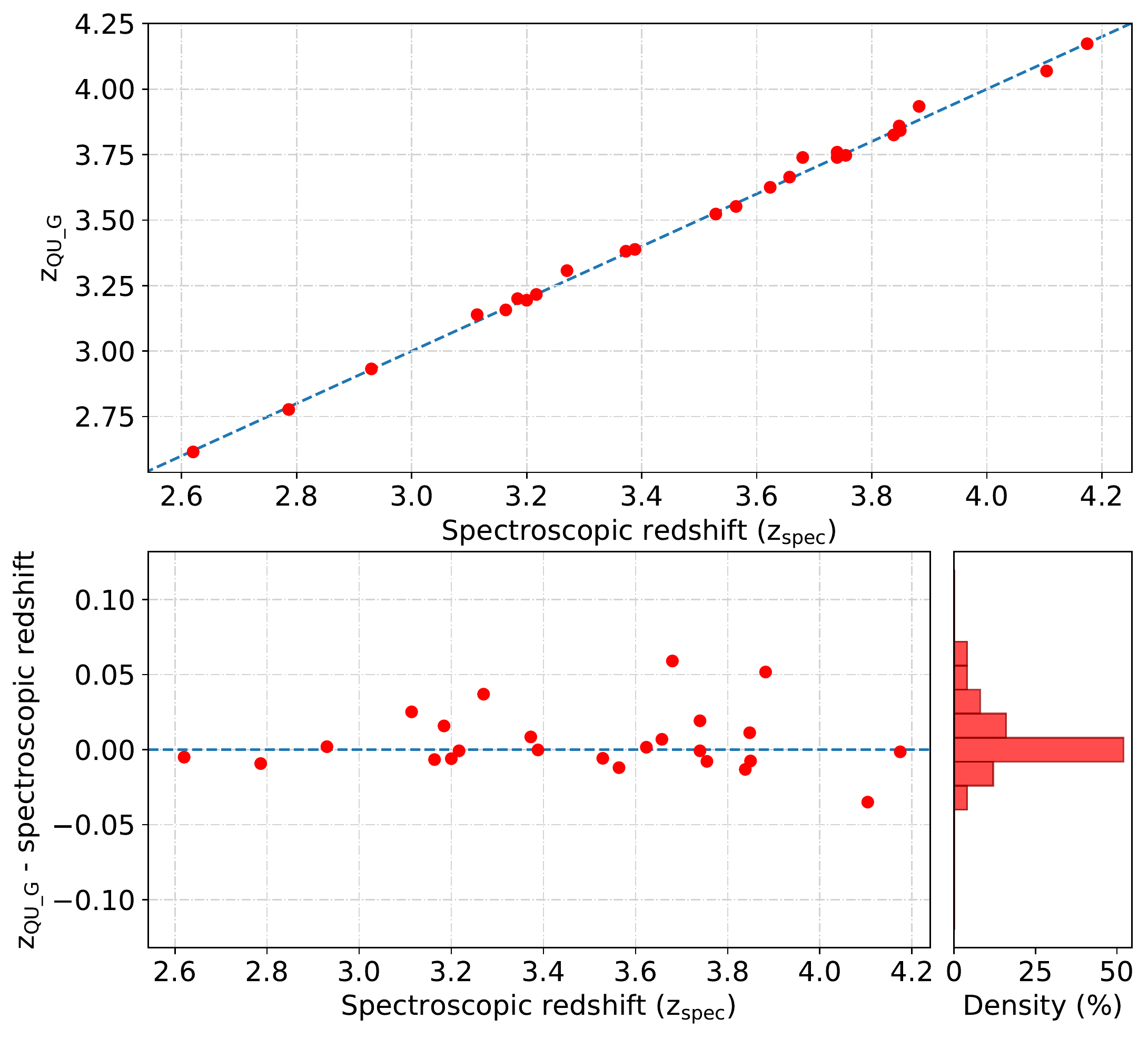}
    \caption{Top Panel: the redshifts determined on the basis the Gaia low-resolution spectra with the procedure described in Sect.\ref{sect:z_from_Gaia} vs. the spectroscopic redshifts for \nObsWithGaiaSpec objects observed spectroscopically.
    Bottom panel: the difference between the redshifts determined on the basis the Gaia low-resolution spectra and the spectroscopic redshifts as a function of the spectroscopic redshfit. 
    }
    \label{Fig:Dz_New}
\end{figure}
\section{Applying the redshift measurement to QUBRICS candidates}
 
 The procedure described in Sect.~\ref{sect:z_from_Gaia}
 has been applied to
 \nCandidatesWithGaiaSpec{}
 quasar candidates produced in Papers III, VI and VII
 of the QUBRICS survey with a Gaia low-resolution spectrum.
 The properties of these \nCandidatesWithGaiaSpec QSO candidates are not significantly different, either in photometric coverage or in magnitude/photometric uncertainty, from the \nKnownQSOHighzWithGaiaSpec{} spectroscopically confirmed QSOs in the QUBRICS database with $\zsp>2.5$, described in Sect.~\ref{sect:z_from_Gaia}.
Therefore, we expect to obtain the same reliability (a test on a limited number of \nObsWithGaiaSpec candidates is carried out in Sect.~\ref{SEC:SpectrFollowUp}).
 
 For \nNewRedshifts objects the procedure produces a redshift estimate of
 sufficient quality (QOP$\ge 2$) to be considered secure.
 They are listed in Tab.~\ref{tab:NewSpec}.
 For \nGaiaSpecNoz objects the absence of spectral features with sufficient SNR
 prevented the determination of a redshift with the requested confidence.
 
 Fig.~\ref{Fig:z_hist_cand} shows the redshift distribution of the \nNewRedshifts 
new identifications. It corresponds to the expected distribution for the QUBRICS candidates (see, for example, Fig.~7 in \citet{Guarneri:2021MNRAS.506.2471G}).
\section{Testing the new redshift measurements with follow-up spectroscopy}
\label{SEC:SpectrFollowUp}

After June 2022, \nObsWithGaiaSpec{}candidates contained in Tab.~\ref{tab:NewSpec} 
(typically chosen on the basis of sky accessibility)
have been observed with follow-up spectroscopy 
at the Telescopio Nazionale Galileo (TNG, La Palma), Las Campanas Observatory
and ESO-NTT using the DOLORES, LDSS-3 (Clay Telescope) and EFOSC-2 spectrographs, respectively. 

Tab.\ref{tab:observations} summarises the observing setups and significant information about each observing run. Figure \ref{fig:exampleSpectra} shows three spectra representative of the three telescopes.

For the 13 candidates observed with the DOLORES instrument, mounted on the Telescopio Nazionale Galileo, exposures have been taken during the AOT45 period, in August 2022, under a proposal with PI F. Guarneri. The LR-B grism (covering a wavelength range between 3600--8000 {\AA} at a
resolution $\sim600$) with a 1" slit aperture was used with an exposure time between 300 and 600 s.

For the 9 targets observed with LDSS-3 at the Clay Telescope, in July and August 2022, observations were obtained on several nights with varied conditions (e.g., bright time, variable weather conditions). The VPH-all grism with the 1" central slit and no blocking filter was employed, covering a wavelength range between 4000--10000 {\AA} with a low resolution of R$\sim$800. Exposure times ranging between 800--1800 s were used, depending on the candidate magnitude and seeing conditions.
\begin{figure}
    \centering
    \includegraphics[width=\columnwidth]{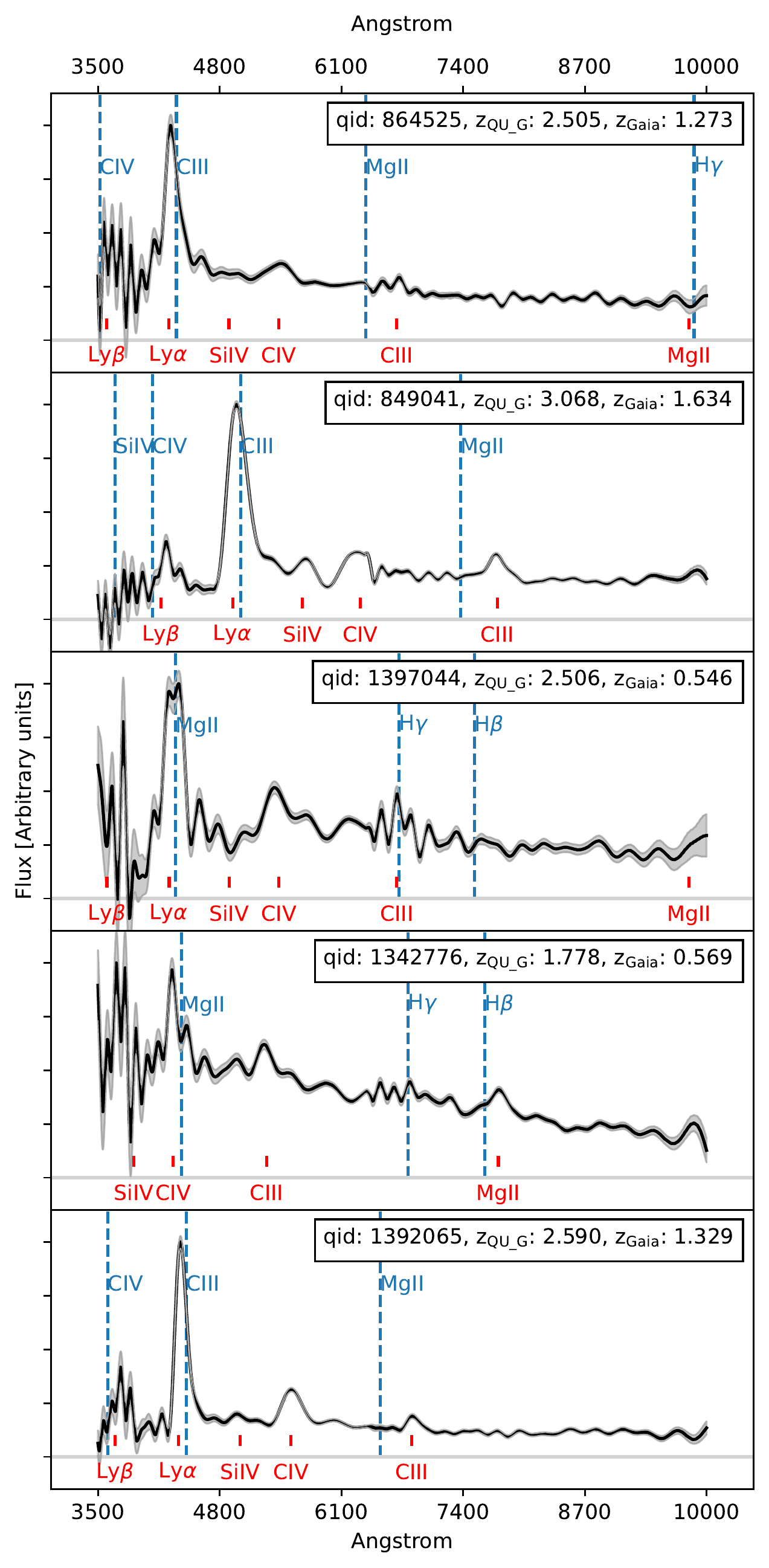}
    \caption{The 5 catastrophic disagreements between QUBRICS' and Gaia's redshifts. The black line shows the Gaia low-resolution spectra with associated error (grey shaded area). Red marks indicate the position of prominent emission lines at the QUBRICS' redshift $\zQUG$, blue dashed lines at the redshift estimated by the Gaia pipeline ($z_\mathrm{Gaia}$).}
    \label{fig:catastrophicFailures}
\end{figure}

Finally, 3 candidates were observed in November 2022, as a part of an observing program at the ESO NTT (PI. F. Guarneri, proposal 110.23WP.001), employing the EFOSC-2 instrument and Grism \#13 (wavelength range $\lambda\sim 3700-9300$ {\AA} and resolution $\sim 1000$), with typical exposure times ranging between 500 and 1200 s.

Data obtained with the DOLORES, LDSS-3 and EFOSC-2 instruments were reduced with a custom pipeline based on MIDAS scripts \citep{MIDAS1988}. Each spectrum has been processed to subtract the bias and normalised by the flat; wavelength calibration is achieved using helium, neon and argon lamps, finding a rms of $\sim0.5${\AA}. Observing conditions have not always been photometric.

All the 25 objects turned out to be QSOs and Fig.~\ref{Fig:Dz_New} shows the good agreement
of the spectroscopic redshifts with the redshifts obtained with the procedure described in Sect.~\ref{sect:z_from_Gaia}: the distribution is characterized by a median $<\Delta z>= -0.0004$
and a standard deviation $\sigma_{\Delta z} = 0.02$, thus confirming the Sect.~\ref{sec:MARZfit} estimates.

\begin{table}
    \centering
    \begin{tabular}{c|c|c|c|c|c}
    \toprule
        \# of objects & Instrument & Telescope & Grism & Slit & Resolution \\
        \midrule
        13  & DOLORES & TNG    & LR-B      & 1"   & 600  \\
        9 & LDSS-3  & Clay   & VPH-all   & 1"   & 800  \\
        3  & EFOSC-2 & NTT    & Grism\#13 & 1.5" & 1000 \\
        \bottomrule
    \end{tabular}
    \caption{Setup for each observing run.}
    \label{tab:observations}
\end{table}

\section{Discussion} 
\label{sec:conclusions}

The tests described in Sect.~\ref{sec:MARZfit} and the spectroscopic follow-up observations described in Sect.~\ref{SEC:SpectrFollowUp}
demonstrate that the procedure described in the present paper produces secure redshifts from the Gaia low-resolution spectra, with an uncertainty of
$\sigma_z \sim 0.02$.

Thus, we can confidently add \nNewRedshifts new spectroscopic redshifts to the QUBRICS database, reducing to \nCandidatesLeft the number of remaining candidates still to be observed.
The large majority ($\sim$80\%) of the \nCandidatesLeft remaining candidates have magnitudes fainter than $i = 18$ and expected redshift below $z = 3$. 
We will continue to follow them up with spectroscopic observations, prioritizing
the construction of complete samples at relatively bright magnitudes ($z>2.5$ and $i<18.0$, $\sim$500 candidates)
and at higher redshift ($z>3$ and $Y<18.5$, $\sim$180 candidates).

A comparison of the redshifts determined in the present work with the redshifts estimated from Gaia \citep{GaiaEDr3:2021A&A...649A...1G} shows a generally good correspondence: out of the \nNewRedshifts redshift determinations of the present work,
1668 also have a redshift estimated by the Gaia collaboration \citep{GaiaEDr3:2021A&A...649A...1G} and in 1663 cases the agreement is within a
$|\Delta z|  \le 0.06 $ with a $\sigma_{\Delta z} = 0.009$. 
However, in 5 cases a catastrophic disagreement occurs. They are shown in Fig. \ref{fig:catastrophicFailures}. We do not have follow-up spectroscopy for these objects,
so only the Gaia low-resolution spectra are shown. It is remarkable that the 5 objects are relatively bright, ${\rm Gaia~ G} \leq 17.63$, and
according to our analysis, based on the typical EW of the emission lines and the continuum slope, in these 5 cases the Gaia pipeline misclassifies some emission line, most notably the Lyman-$\alpha$ or \ion{C}{IV} lines as a \ion{C}{III$_{1909}$}, as shown in Fig.~\ref{fig:catastrophicFailures}.

\begin{figure}
	\includegraphics[width=\columnwidth]{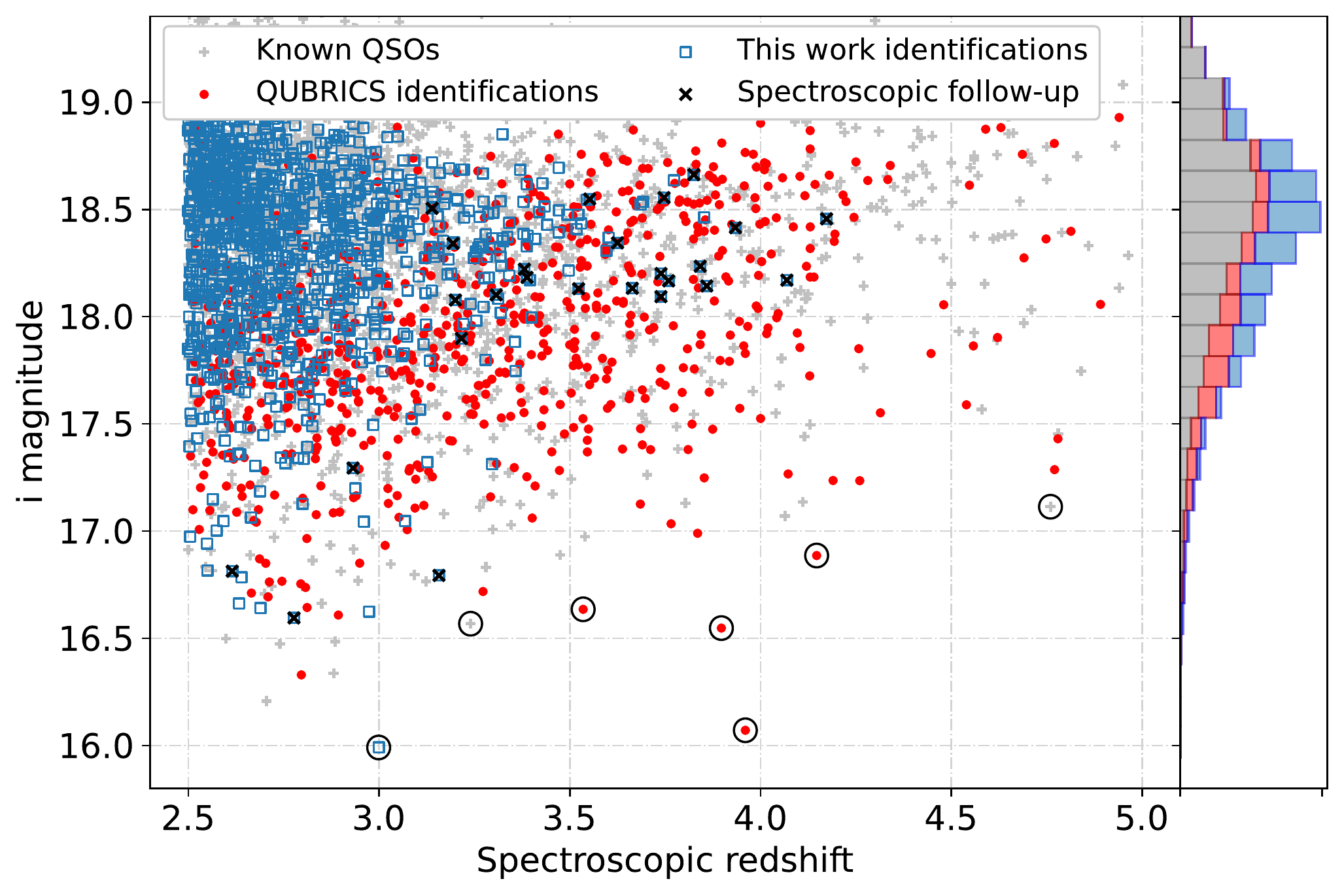}
    \caption{Known QSO in the Southern Hemisphere from the literature (grey '+'), identifications from QUBRICS before this work (red dots) and identifications from this work (blue, empty squares). Black crosses ('x') mark objects with a redshift determination from Gaia confirmed through spectroscopic follow-up (see last three columns in Tab. \ref{tab:NewSpec}). The histogram on the right highlights the significant contribution of QUBRICS in the search for bright beacons in the south. Circled QSOs represent the new Golden Sample for the Sandage Test, updated with respect to Paper II (see Sect.~\ref{sec:Sandage}).
    }
    \label{fig:ivsz}
\end{figure}

Fig.~\ref{fig:ivsz} shows the new diagram with the known quasars at $z>2.5$ in the Southern hemisphere vs. their $i$ magnitude.
The contribution of the QUBRICS survey, especially at the brighter magnitudes, is apparent. 
\subsection{The Sandage Test of the redshift drift}
\label{sec:Sandage}
In Paper II we showed that, thanks to the new bright QSO identifications at high redshift, the total time required to carry out the Sandage Test of the cosmologic redshift drift \citep{Sandage62} could be reduced  to less than 2500 hours of observations with an ANDES-like \citep{ANDES_MArconi2022} spectrograph at the ELT. In particular we envisaged the use of the fiber-fed VIS-BLUE (UBV) and VISRED (RIZ) modules of ANDES to obtain in natural seeing conditions high-resolution ($R \sim 100,000$), high-fidelity spectra of 30 targets, with a temporal separation of 25 years.
We are now in the position to update these estimates by defining a new "Golden Sample":  two main refinements are the inclusion of new bright QSOs discovered with QUBRICS and a realistic observing strategy \citep{Dong2022}
with a reduced number of targets, but still sufficient
to have at least one suitable QSO available for observations from the ELT site at any time of the year. 

In the following we adopt the same assumptions as in Paper II (updated from \cite{Liske+08:2008MNRAS.386.1192L}), namely the uncertainty in the radial velocity measurement, $\sv$:
\begin{equation}
\sv = g \times 1.35 
\left(\frac{{S/N}}{3350}\right)^{-1} 
\left(\frac{1 + \zqso}{5}\right)^{-\gamma} 
\left(\frac{\nqso}{30}\right)^{-0.5}
{\rm cm/s}
\label{sveqmne}
\end{equation}
where the symbol `S/N' refers to the total S/N per $0.0125$~\AA\ pixel per
object accumulated over all observations, $\nqso$ is the number of QSOs in the sample, $\zqso$ is the redshift of the QSO, the $\gamma$ exponent is $1.7$ for $\zqso \leq 4$ and 0.9 above. The form factor $g$ is equal to $1$ if all the targets are
observed twice, at the beginning and at the end of the experiment, and becomes larger if the measurements are distributed in time, reaching $1.7$ for a uniform distribution.
The S/N per pixel for photon-noise limited observations can be written as:
\begin{equation}
\label{sn}
{S/N} = 650 \left[ \frac{Z_X}{Z_r} \; 10^{0.4 (16 -
    m_X)} \; \left( \frac{D}{39 {\rm m}} \right)^2 \; \frac{\ti}{\rm 10 h} \; 
    \frac{\epsilon}{0.25} \right]^\frac{1}{2}
\end{equation}
where $D$, $\ti$ and $\epsilon$ are the telescope diameter, total
integration time and total efficiency, $Z_X$ and $m_X$ are the zeropoint
and apparent magnitude of the source in the $X$-band, respectively,
and $Z_r = (8.88 \times 10^{10})$~s$^{-1}$~m$^{-2}$~$\mu$m$^{-1}$ is
the AB zeropoint for an effective wavelength of
$6170$~\AA\ (corresponding to the SDSS $r$-band). The normalisation of the above equation assumes a pixel
size of $0.0125$~\AA\ and a central
obscuration of the telescope's primary collecting area of $10$ per
cent.
\begin{figure}
 \centering
	\includegraphics[width=\columnwidth]{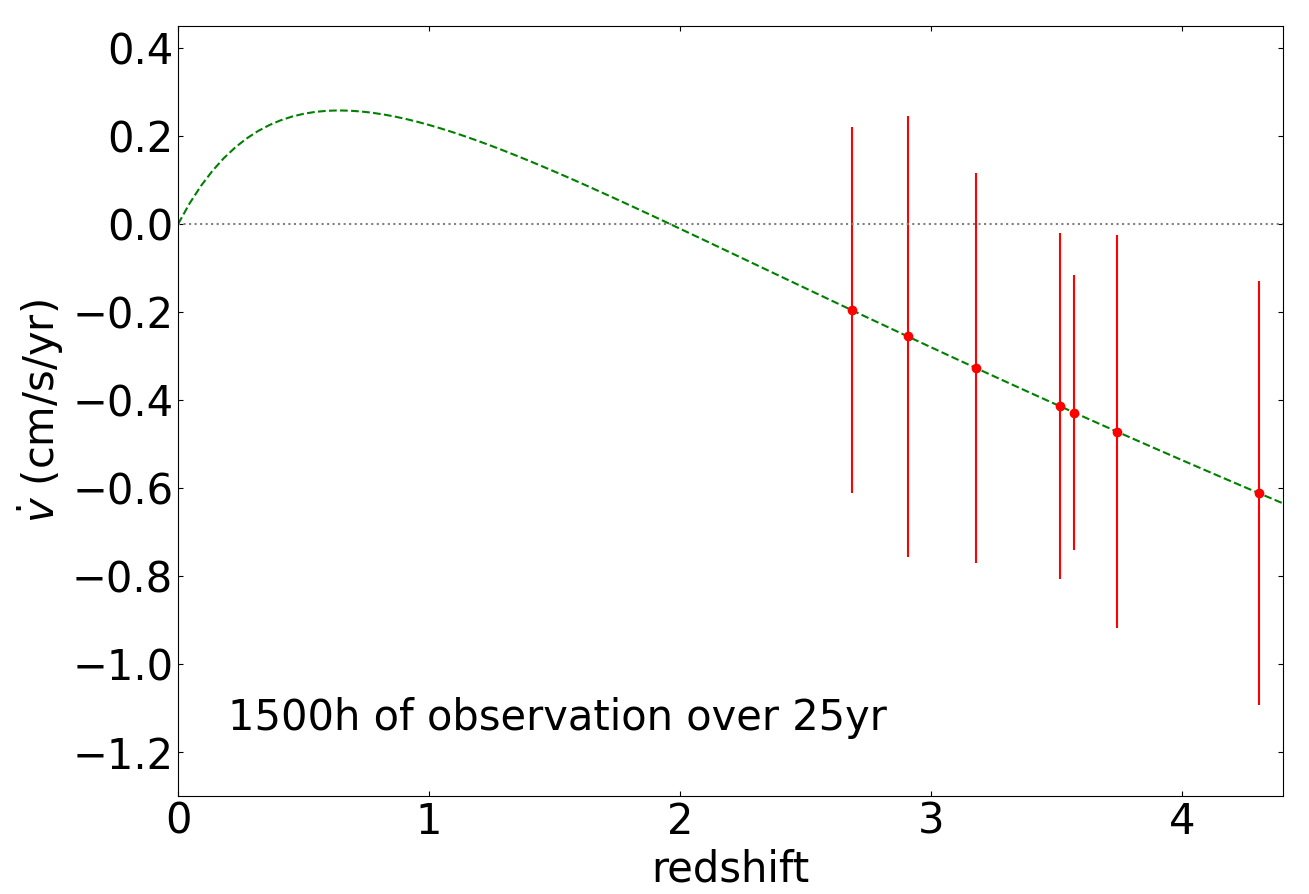}
 	\includegraphics[width=\columnwidth]{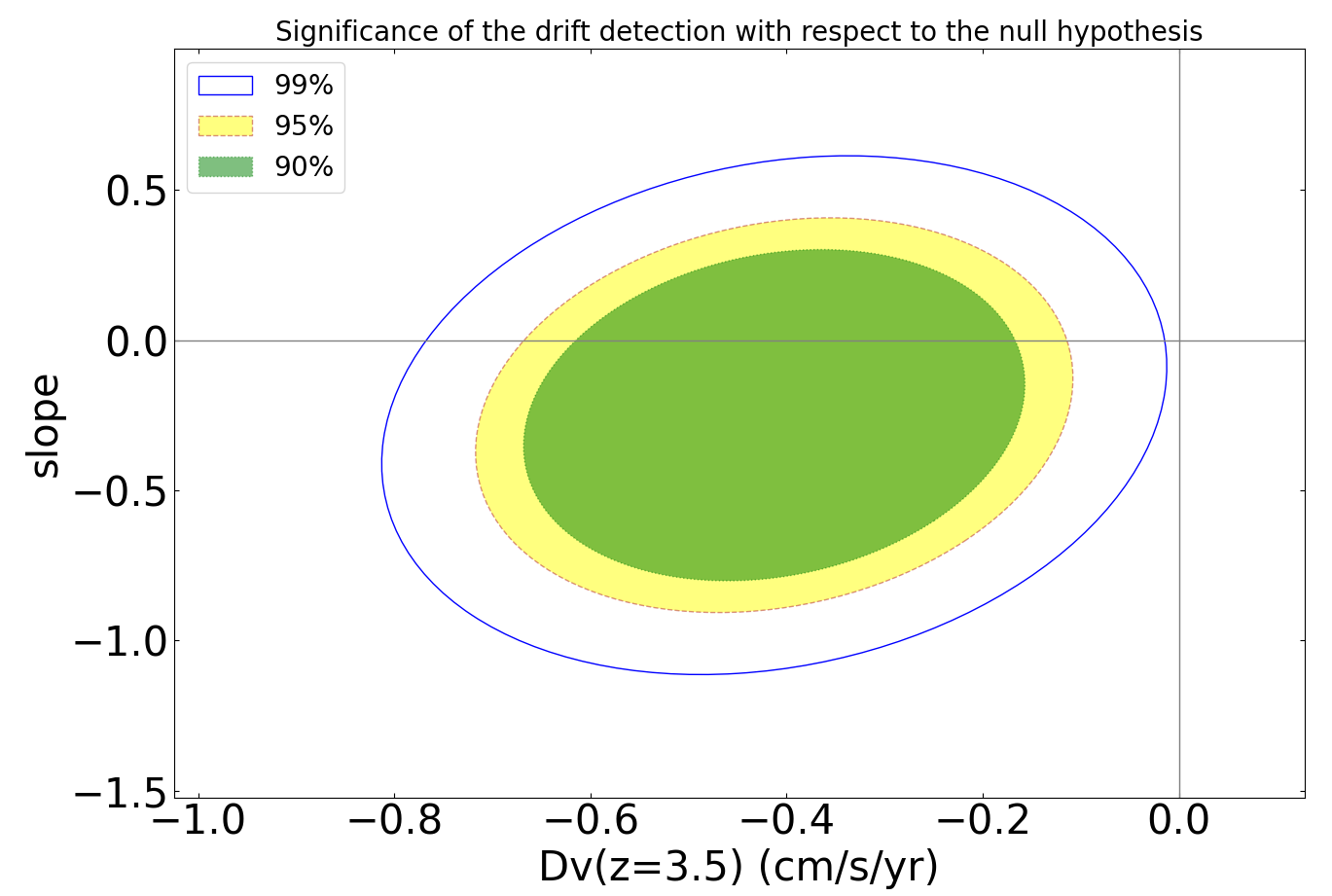}
    \caption{Simulated measurement of the redshift drift with the seven QSOs of the Golden Sample of Tab.~\ref{tab:gold}. 3000 random realizations have been fitted each with a linear trend, $\dot v = {\rm slope} \times (z-3.5) + \Delta v(z=3.5)$. In the upper panel the red bars represent the expected standard deviation and the green dotted line shows the expected signal of the redshift drift in a PLANCK18 cosmology.
    The green, yellow and blue-line-encircled regions in the lower panel show the $90\%$ and $95\%$ and $99\%$ confidence regions, respectively, for the measured $\dot v$.
    }
    \label{fig:Sandage}
\end{figure}
\begin{figure}
 \centering
	\includegraphics[width=\columnwidth]{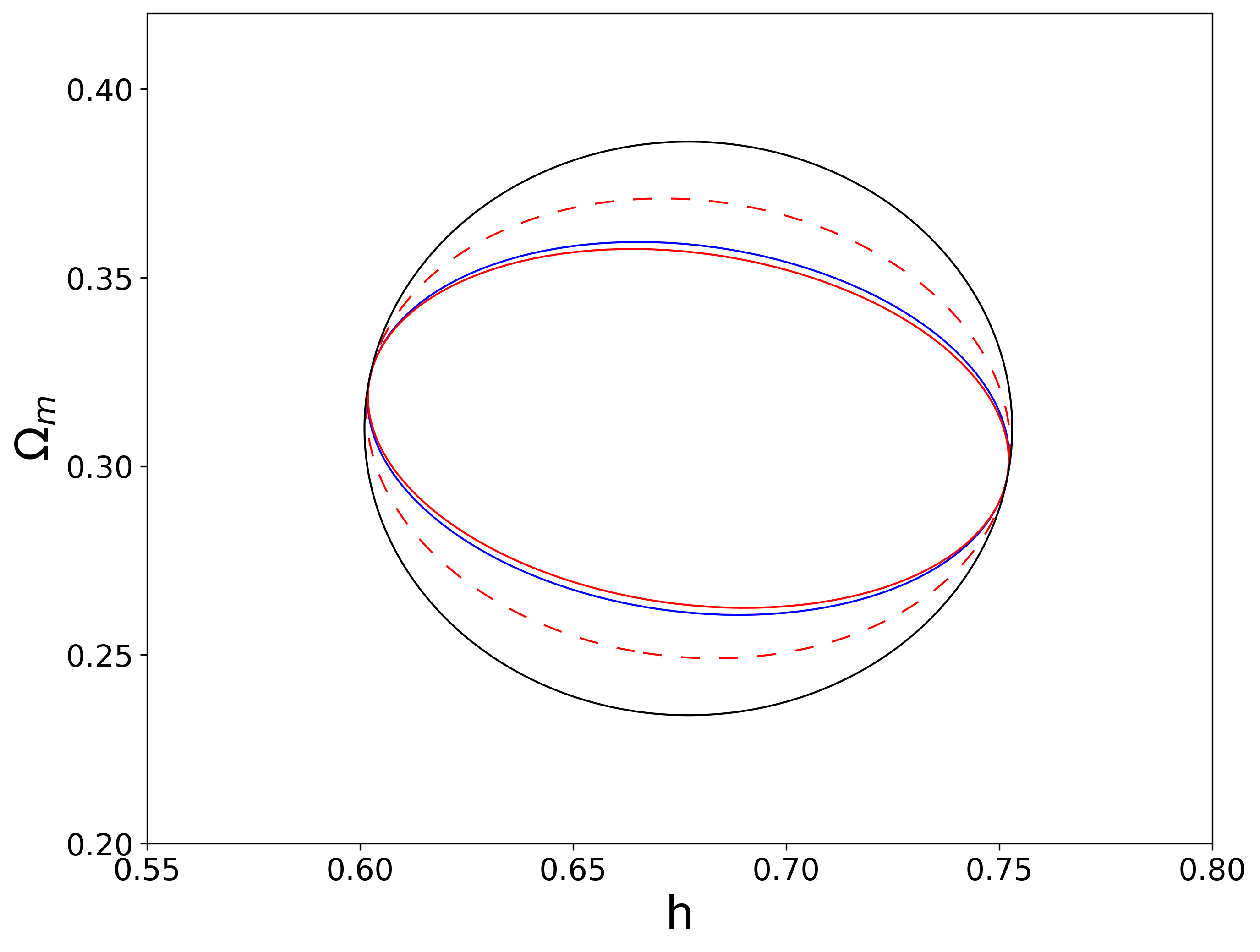}
    \caption{One-sigma confidence ellipses from our Fisher Matrix analysis of the expected data of the Golden Sample obtained  with 1500 h of observation at the ELT over 25 years. In solid red we have the Golden Sample observed with a form factor g=1 (all the targets are observed twice, at the beginning and at the end of the experiment) and dash red for g=1.7 (uniform distribution of the observations). In blue we show the previous result reported in Paper II for a Sample of 30 QSO,  with 2500 hours of observation and a form factor g=1. The black line is the one-sigma constraint from the priors only.
    }
    \label{fig:FisherQUBRICS}
\end{figure}
\begin{figure}
	\includegraphics[width=\columnwidth]{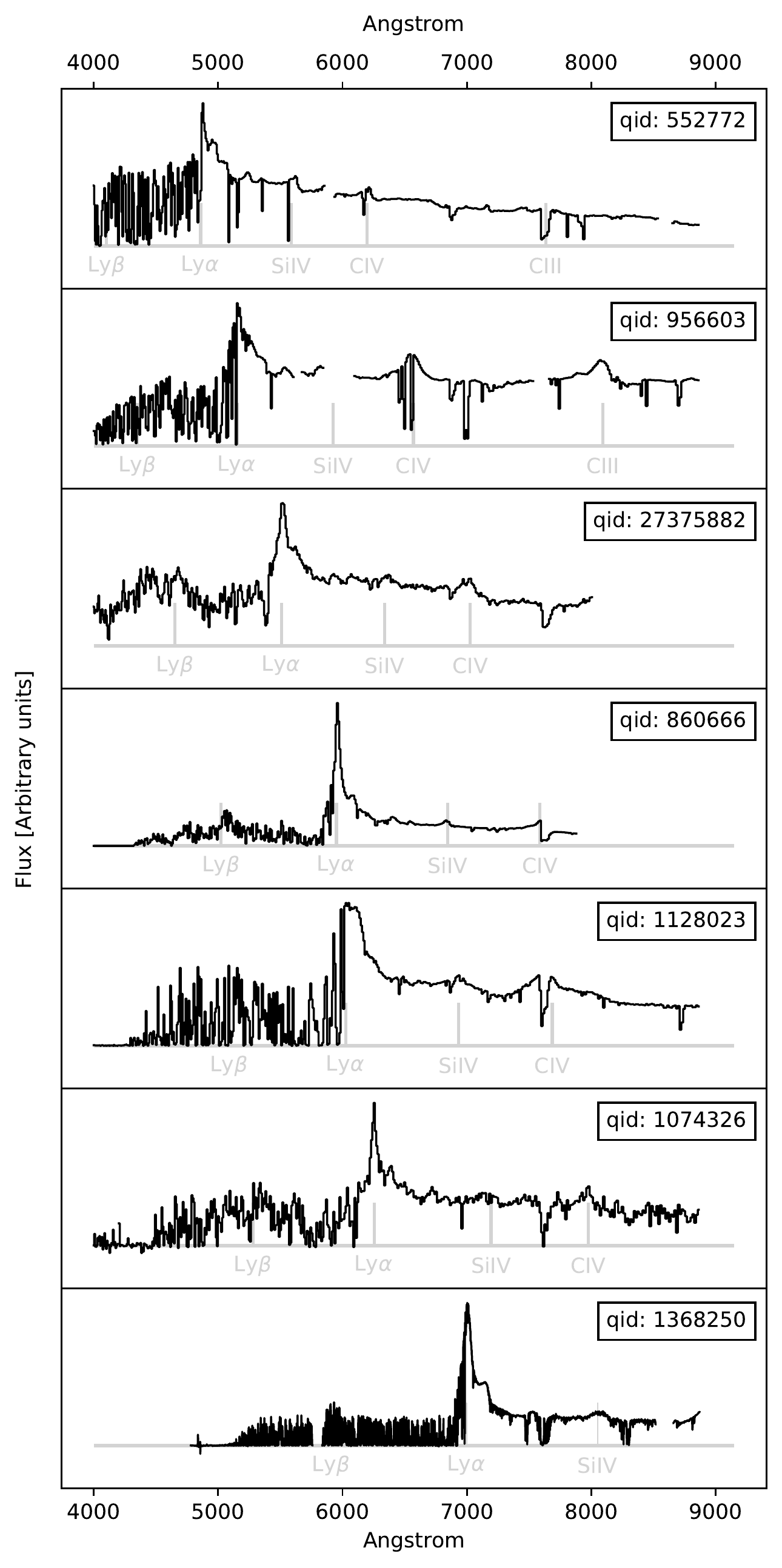}
    \caption{Spectra of the seven QSOs presented in Tab. \ref{tab:gold}. Spectra have been taken from the QUBRICS database (qid 860666, 1128023, 1074326 and 27375882), the UVES SQUAD dataset \citep[qid 956603,][]{MMurphy:squad}, the SpecDB database \citep[qid 552772 and 1368250,][]{specDB}. For visualization purposes, all spectra have been re-sampled to 500 km s$^{-1}$, with the exception of 1368250, sampled at 50 km s$^{-1}$.
    }
    \label{fig:GoldenQSOPlot}
\end{figure}

With respect to Paper II new brighter quasars (e.g. qid 1128023) have been found and we have reduced the Golden Sample to the seven brightest QSOs
in the range $2.9 < z < 4.8$. 
The seven QSOs cover a wide range of right ascension and this favors observability/scheduling.
We are also assuming a fixed allocation of $1500/7 = 214$h of observation for each QSO (and not a variable amount of time required to reach a $\sv$ = 22.8 \cms as in Paper II). 
In this way, assuming the most optimistic form factor $g = 1$ (which holds if all the targets are observed twice, at the beginning and at the end of the experiment),
for a total of 1500 hours of observations in 25 years, the redshift drift is expected to be detected with a confidence level of the order of 99\% (as shown in Fig.~\ref{fig:Sandage}).
Besides, the combination of the measurement of the redshift drift with the QSOs Lyman Forest,
at $z \magcir 2.5$, will be plausibly complemented at $z \mincir 2.5$ 
with radio observations by the SKAO, CHIME and FAST \citep{Moresco2022}, providing an extended redshift leverage of the drift and significantly shortening the observing time required.
The goal of the Sandage test is to provide a direct, real-time mapping of the expansion rate of the universe, independent of assumptions on gravity, geometry or clustering, so the precision of the determination of cosmological parameters is not its main focus. Nevertheless, recent theoretical studies \citep{Martins2016,Alves2019,Esteves2021} have uncovered important synergies with other cosmological probes, including the characterization of the physical properties of dark energy. 
In this respect it is of interest to mention that, 
assuming a flat $\Lambda$CDM cosmological model with the \cite{Planck18} best-fit cosmological parameters, $\Omega_\mathrm{m}=0.31\ $ and $H_0 = h ~ 100 =67.7\  \mathrm{kms^{-1}Mpc^{-1}}$, and external
priors for both of these parameters of $\sigma_{h}=0. 05$ and
$\sigma_{\Omega_m}=0.05$, a Fisher Matrix analysis leads to a one-sigma
uncertainty for $\Omega_m$ of 0.03 when the form factor is g=1 and of 0.04
with g=1.7, while the $h$ constraint recovers the prior. Fig.~ \ref{fig:FisherQUBRICS} compares these one-sigma constraints with the ones from the previous version of
the Golden Sample in Paper II. A small gain in the matter
density constraint, in spite of the  1000 h reduction of the observing time, is clearly visible (comparing for consistency the two g=1 cases). 

The spectra of the seven QSOs in the Golden Sample have also been checked for the suitability of their absorption spectra for the redshift drift measurement
(indeed, an eighth QSO has been excluded, due to the presence of BAL features) and are shown in Fig.~\ref{fig:GoldenQSOPlot}.

The brightness of the QSOs of the Golden Sample makes the beginning of observations of this type conceivable in advance of the realization of the ELT, for example with the super-stable ESPRESSO spectrograph \citep{ESPRESSO_Pepe2021} at the ESO VLT and a time investment of 3-4 nights per month. Indeed, a pilot program, 110.247Q.001 "An ESPRESSO Redshift Drift Experiment", started at ESO in 2022.
\begin{table*}
    \centering
    \begin{tabular}{r|c|c|c|c|c|c|c|}
\toprule
   qid      & RA            & DEC     & $r_{\rm psf}$ & $i_{\rm psf}$ & $G_{\rm GAIA}$ & redshift  &  $\sv$ in 214h\\
             & J2000         & J2000         & (mag)     & (mag)    &  (mag) &       &           (\cms)  \\
\midrule
% 1352951 & 00:18:29.95 & +03:19:03.3 &  16.467 & 16.327 & 16.6646 & 2.777 & - \\
  956603 & 00:41:31.44 & -49:36:11.7 &  16.675 & 16.568 & 16.6981 & 3.240 & 12.5 \\
% 1472544 & 01:03:11.27 & +13:16:17.8 &  16.350 & 16.207 & 16.2785 & 2.705 & - \\
%  953138 & 01:56:44.71 & -69:22:16.2 &  16.633 & 16.329 & 16.5897 & 2.796 & - \\ 
1128023  & 05:29:15.81 & -43:51:52.1 &  16.264 & 16.071 & 16.3452 & 3.960 & ~7.8 \\
1074326  & 05:48:03.20 & -48:48:13.1 &  17.215 & 16.886 & 17.2414 & 4.147 & 11.2 \\ 
 1368250 & 14:51:47.05 & -15:12:20.2 &  18.884 & 17.113 & 18.0754 & 4.760 & 12.1 \\
% 1338905 & 16:08:43.90 & +07:15:08.7 &  16.619  & 16.336 & 16.5227 & 2.881 & - \\
  860666 & 21:25:40.97 & -17:19:51.4 & 16.644 & 16.548 & 16.8730 & 3.897 & ~9.8 \\ 
%  875768 & 21:57:28.23 & -36:02:15.2 & 18.821 & 17.286 & 18.2650 & 4.685 & 17.2 \\
  552772 & 22:47:08.93 & -60:15:45.3 & 16.135 & 15.991 & 16.0895 & 2.999 & 10.4 \\
27375882 & 22:54:51.33 & -05:29:24.0 & 16.641 & 16.635 & 16.7257 & 3.535 & 11.1 \\
\bottomrule
\end{tabular}
\caption{The golden sample of Southern QSOs for the Sandage Test}
\label{tab:gold}
\end{table*}

%
%The last numbered section should briefly summarise what has been done, and describe
%the final conclusions which the authors draw from their work.
%
\section*{Acknowledgements}
We thank an anonymous referee for comments that prompted a significant improvement of this paper.
This work was financed by Portuguese funds through FCT - Funda\c{c}ao para a Ci\^encia e a Tecnologia in the framework of the project 2022.04048.PTDC. CJM also acknowledges FCT and POCH/FSE (EC) support through Investigador FCT Contract 2021.01214.CEECIND/CP1658/CT0001.
We thank Societ\`a Astronomica Italiana (SAIt), Ennio Poretti, Gloria Andreuzzi and Marco Pedani for the observation support at TNG. Part of the observations discussed in this work are based on observations made with the Italian Telescopio Nazionale Galileo (TNG) operated on the island of La Palma by the Fundacion Galileo Galilei of the INAF (Istituto Nazionale di Astrofisica) at the Spanish Observatorio del Roque de los Muchachos of the Instituto de Astrofisica de Canarias.

The national facility capability for SkyMapper has been funded through ARC LIEF grant LE130100104 from the Australian Research Council, awarded to the University of Sydney, the Australian National University, Swinburne University of Technology, the University of Queensland, the University of Western Australia, the University of Melbourne, Curtin University of Technology, Monash University and the Australian Astronomical Observatory. SkyMapper is owned and operated by the Australian National University’s Research School of Astronomy and Astrophysics. The survey data have been processed and provided by the SkyMapper Team at ANU. The SkyMapper node of the All-Sky Virtual Observatory (ASVO) is hosted at the National Computational Infrastructure (NCI). Development and support the SkyMapper node of the ASVO has been funded in part by Astronomy Australia Limited (AAL) and the Australian Government through the Commonwealth’s Education Investment Fund (EIF) and National Collaborative Research Infrastructure Strategy (NCRIS), particularly the National eResearch Collaboration Toolsand Resources (NeCTAR) and the Australian National Data Service Projects (ANDS).

This work has made use of data from the European Space Agency (ESA) mission Gaia (https://www.cosmos.esa.int/gaia), processed by the Gaia Data Processing and Analysis Consortium (DPAC, https://www.cosmos.esa.int/web/gaia/dpac/consortium). Funding for the DPAC has been provided by national institutions, in particular the institutions participating in the Gaia Multilateral Agreement. This job has made use of the Python package GaiaXPy, developed and maintained by members of the Gaia Data Processing and Analysis Consortium (DPAC), and in particular, Coordination Unit 5 (CU5), and the Data Processing Centre located at the Institute of Astronomy, Cambridge, UK (DPCI).

This publication makes use of data products from the Two Micron All Sky Survey, which is a joint project of the University of Massachusetts and the Infrared Processing and Analysis Center/California Institute of Technology, funded by the National Aeronautics and Space Administration and the National Science Foundation. 

This publication makes use of data products from the Wide-field Infrared Survey Explorer, which is a joint project of the University of California, Los Angeles, and the Jet Propulsion Laboratory/California Institute of Technology, funded by the National Aeronautics and Space Administration. 

%Based on data obtained from the ESO Science Archive Facility with DOI(s): https://doi.org/10.18727/archive/71

This paper includes data gathered with the 6.5 meter Magellan Telescopes located at Las Campanas Observatory, Chile.

%%%%%%%%%%%%%%%%%%%%%%%%%%%%%%%%%%%%%%%%%%%%%%%%%%
\section*{Data Availability}
The data underlying this article will be shared on reasonable request to the corresponding author.



\bibliographystyle{mnras}
\bibliography{00_QUBRICS_GAIA} % if your bibtex file is called example.bib

\begin{thebibliography}{}
\makeatletter
\relax
\def\mn@urlcharsother{\let\do\@makeother \do\$\do\&\do\#\do\^\do\_\do\%\do\~}
\def\mn@doi{\begingroup\mn@urlcharsother \@ifnextchar [ {\mn@doi@}
  {\mn@doi@[]}}
\def\mn@doi@[#1]#2{\def\@tempa{#1}\ifx\@tempa\@empty \href
  {http://dx.doi.org/#2} {doi:#2}\else \href {http://dx.doi.org/#2} {#1}\fi
  \endgroup}
\def\mn@eprint#1#2{\mn@eprint@#1:#2::\@nil}
\def\mn@eprint@arXiv#1{\href {http://arxiv.org/abs/#1} {{\tt arXiv:#1}}}
\def\mn@eprint@dblp#1{\href {http://dblp.uni-trier.de/rec/bibtex/#1.xml}
  {dblp:#1}}
\def\mn@eprint@#1:#2:#3:#4\@nil{\def\@tempa {#1}\def\@tempb {#2}\def\@tempc
  {#3}\ifx \@tempc \@empty \let \@tempc \@tempb \let \@tempb \@tempa \fi \ifx
  \@tempb \@empty \def\@tempb {arXiv}\fi \@ifundefined
  {mn@eprint@\@tempb}{\@tempb:\@tempc}{\expandafter \expandafter \csname
  mn@eprint@\@tempb\endcsname \expandafter{\@tempc}}}

\bibitem[\protect\citeauthoryear{{Alves}, {Leite}, {Martins}, {Matos}  \&
  {Silva}}{{Alves} et~al.}{2019}]{Alves2019}
{Alves} C.~S.,  {Leite} A.~C.~O.,  {Martins} C.~J.~A.~P.,  {Matos} J.~G.~B.,
  {Silva} T.~A.,  2019, \mn@doi [\mnras] {10.1093/mnras/stz1934}, \href
  {https://ui.adsabs.harvard.edu/abs/2019MNRAS.488.3607A} {488, 3607}

\bibitem[\protect\citeauthoryear{Anderson}{Anderson}{2003}]{ref:CCA}
Anderson T.~W.,  2003, An Introduction to Multivariate Statistical Analysis, 3
  edn.
Wiley series in probability and mathematical statistics, Wiley, \url
  {https://www.wiley.com/en-us/An+Introduction+to+Multivariate+Statistical+Analysis%2C+3rd+Edition-p-9780471360919}

\bibitem[\protect\citeauthoryear{{Banse}, {Ponz}, {Ounnas}, {Grosbol}  \&
  {Warmels}}{{Banse} et~al.}{1988}]{MIDAS1988}
{Banse} K.,  {Ponz} D.,  {Ounnas} C.,  {Grosbol} P.,   {Warmels} R.,  1988, in
  Instrumentation for Ground-Based Optical Astronomy. p.~431

\bibitem[\protect\citeauthoryear{{Boutsia} et~al.,}{{Boutsia}
  et~al.}{2020}]{Boutsia2020}
{Boutsia} K.,  et~al., 2020, \mn@doi [\apjs] {10.3847/1538-4365/abafc1}, \href
  {https://ui.adsabs.harvard.edu/abs/2020ApJS..250...26B} {250, 26}

\bibitem[\protect\citeauthoryear{{Boutsia} et~al.,}{{Boutsia}
  et~al.}{2021}]{LF_Boutsia:2021ApJ...912..111B}
{Boutsia} K.,  et~al., 2021, \mn@doi [\apj] {10.3847/1538-4357/abedb5}, \href
  {https://ui.adsabs.harvard.edu/abs/2021ApJ...912..111B} {912, 111}

\bibitem[\protect\citeauthoryear{{Calderone}, {Nicastro}, {Ghisellini},
  {Dotti}, {Sbarrato}, {Shankar}  \& {Colpi}}{{Calderone}
  et~al.}{2017}]{QSOFITS_2017}
{Calderone} G.,  {Nicastro} L.,  {Ghisellini} G.,  {Dotti} M.,  {Sbarrato} T.,
  {Shankar} F.,   {Colpi} M.,  2017, \mn@doi [\mnras] {10.1093/mnras/stx2239},
  \href {https://ui.adsabs.harvard.edu/abs/2017MNRAS.472.4051C} {472, 4051}

\bibitem[\protect\citeauthoryear{{Calderone} et~al.,}{{Calderone}
  et~al.}{2019}]{Calderone19:2019ApJ...887..268C}
{Calderone} G.,  et~al., 2019, \mn@doi [\apj] {10.3847/1538-4357/ab510a}, \href
  {https://ui.adsabs.harvard.edu/abs/2019ApJ...887..268C} {887, 268}

\bibitem[\protect\citeauthoryear{{Chambers} et~al.,}{{Chambers}
  et~al.}{2016}]{PanSTARRS:Chambers_2016}
{Chambers} K.~C.,  et~al., 2016, arXiv e-prints, \href
  {https://ui.adsabs.harvard.edu/abs/2016arXiv161205560C} {p. arXiv:1612.05560}

\bibitem[\protect\citeauthoryear{Chen \& Guestrin}{Chen \&
  Guestrin}{2016}]{XGBoost2016}
Chen T.,  Guestrin C.,  2016, in Proceedings of the 22nd ACM SIGKDD
  International Conference on Knowledge Discovery and Data Mining. KDD '16.
ACM, New York, NY, USA, pp 785--794, \mn@doi{10.1145/2939672.2939785}, \url
  {http://doi.acm.org/10.1145/2939672.2939785}

\bibitem[\protect\citeauthoryear{{Colless} et~al.,}{{Colless}
  et~al.}{2001}]{2df:2001MNRAS.328.1039C}
{Colless} M.,  et~al., 2001, \mn@doi [\mnras]
  {10.1046/j.1365-8711.2001.04902.x}, \href
  {https://ui.adsabs.harvard.edu/abs/2001MNRAS.328.1039C} {328, 1039}

\bibitem[\protect\citeauthoryear{{Cupani} et~al.,}{{Cupani}
  et~al.}{2022}]{ref:cupaniFeLoBALs}
{Cupani} G.,  et~al., 2022, \mn@doi [\mnras] {10.1093/mnras/stab3562}, \href
  {https://ui.adsabs.harvard.edu/abs/2022MNRAS.510.2509C} {510, 2509}

\bibitem[\protect\citeauthoryear{{De Angeli} et~al.,}{{De Angeli}
  et~al.}{2022}]{GaiaSp2022}
{De Angeli} F.,  et~al., 2022, arXiv e-prints, \href
  {https://ui.adsabs.harvard.edu/abs/2022arXiv220606143D} {p. arXiv:2206.06143}

\bibitem[\protect\citeauthoryear{{Dong}, {Gonzalez}, {Eikenberry}, {Jeram},
  {Likamonsavad}, {Liske}, {Stelter}  \& {Townsend}}{{Dong}
  et~al.}{2022}]{Dong2022}
{Dong} C.,  {Gonzalez} A.,  {Eikenberry} S.,  {Jeram} S.,  {Likamonsavad} M.,
  {Liske} J.,  {Stelter} D.,   {Townsend} A.,  2022, \mn@doi [\mnras]
  {10.1093/mnras/stac1702}, \href
  {https://ui.adsabs.harvard.edu/abs/2022MNRAS.514.5493D} {514, 5493}

\bibitem[\protect\citeauthoryear{{Esteves}, {Martins}, {Pereira}  \&
  {Alves}}{{Esteves} et~al.}{2021}]{Esteves2021}
{Esteves} J.,  {Martins} C.~J.~A.~P.,  {Pereira} B.~G.,   {Alves} C.~S.,  2021,
  \mn@doi [\mnras] {10.1093/mnrasl/slab102}, \href
  {https://ui.adsabs.harvard.edu/abs/2021MNRAS.508L..53E} {508, L53}

\bibitem[\protect\citeauthoryear{{Flesch}}{{Flesch}}{2013}]{FleschVeronCorrections}
{Flesch} E.,  2013, \mn@doi [\pasa] {10.1017/pasa.2012.004}, \href
  {https://ui.adsabs.harvard.edu/abs/2013PASA...30....4F} {30, e004}

\bibitem[\protect\citeauthoryear{{Fontanot} et~al.,}{{Fontanot}
  et~al.}{2023}]{Fontanot2023}
{Fontanot} F.,  et~al., 2023, \mn@doi [\mnras] {10.1093/mnras/stad189}, \href
  {https://ui.adsabs.harvard.edu/abs/2023MNRAS.520..740F} {520, 740}

\bibitem[\protect\citeauthoryear{{Gaia Collaboration} et~al.,}{{Gaia
  Collaboration} et~al.}{2021}]{GaiaEDr3:2021A&A...649A...1G}
{Gaia Collaboration} et~al., 2021, \mn@doi [\aap]
  {10.1051/0004-6361/202039657}, \href
  {https://ui.adsabs.harvard.edu/abs/2021A&A...649A...1G} {649, A1}

\bibitem[\protect\citeauthoryear{{Grazian} et~al.,}{{Grazian}
  et~al.}{2022}]{LF_Hz_Grazian:2021arXiv211013736G}
{Grazian} A.,  et~al., 2022, \mn@doi [\apj] {10.3847/1538-4357/ac33a4}, \href
  {https://ui.adsabs.harvard.edu/abs/2022ApJ...924...62G} {924, 62}

\bibitem[\protect\citeauthoryear{{Guarneri}, {Calderone}, {Cristiani},
  {Fontanot}, {Boutsia}, {Cupani}, {Grazian}  \& {D'Odorico}}{{Guarneri}
  et~al.}{2021}]{Guarneri:2021MNRAS.506.2471G}
{Guarneri} F.,  {Calderone} G.,  {Cristiani} S.,  {Fontanot} F.,  {Boutsia} K.,
   {Cupani} G.,  {Grazian} A.,   {D'Odorico} V.,  2021, \mn@doi [\mnras]
  {10.1093/mnras/stab1867}, \href
  {https://ui.adsabs.harvard.edu/abs/2021MNRAS.506.2471G} {506, 2471}

\bibitem[\protect\citeauthoryear{{Guarneri} et~al.,}{{Guarneri}
  et~al.}{2022}]{Guarneri2022}
{Guarneri} F.,  et~al., 2022, \mn@doi [\mnras] {10.1093/mnras/stac2733}, \href
  {https://ui.adsabs.harvard.edu/abs/2022MNRAS.517.2436G} {517, 2436}

\bibitem[\protect\citeauthoryear{{Hinton}, {Davis}, {Lidman}, {Glazebrook}  \&
  {Lewis}}{{Hinton} et~al.}{2016}]{MARZ_2016}
{Hinton} S.~R.,  {Davis} T.~M.,  {Lidman} C.,  {Glazebrook} K.,   {Lewis}
  G.~F.,  2016, \mn@doi [Astronomy and Computing]
  {10.1016/j.ascom.2016.03.001}, \href
  {https://ui.adsabs.harvard.edu/abs/2016A&C....15...61H} {15, 61}

\bibitem[\protect\citeauthoryear{{Inoue}, {Shimizu}, {Iwata}  \&
  {Tanaka}}{{Inoue} et~al.}{2014}]{Inoue2014}
{Inoue} A.~K.,  {Shimizu} I.,  {Iwata} I.,   {Tanaka} M.,  2014, \mn@doi
  [\mnras] {10.1093/mnras/stu936}, \href
  {https://ui.adsabs.harvard.edu/abs/2014MNRAS.442.1805I} {442, 1805}

\bibitem[\protect\citeauthoryear{{Jones} et~al.,}{{Jones}
  et~al.}{2009}]{6df:2009MNRAS.399..683J}
{Jones} D.~H.,  et~al., 2009, \mn@doi [\mnras]
  {10.1111/j.1365-2966.2009.15338.x}, \href
  {https://ui.adsabs.harvard.edu/abs/2009MNRAS.399..683J} {399, 683}

\bibitem[\protect\citeauthoryear{{Kubota} \& {Done}}{{Kubota} \&
  {Done}}{2018}]{PYAGN_2018}
{Kubota} A.,  {Done} C.,  2018, \mn@doi [\mnras] {10.1093/mnras/sty1890}, \href
  {https://ui.adsabs.harvard.edu/abs/2018MNRAS.480.1247K} {480, 1247}

\bibitem[\protect\citeauthoryear{{Laor} \& {Netzer}}{{Laor} \&
  {Netzer}}{1989}]{Laor1989}
{Laor} A.,  {Netzer} H.,  1989, \mn@doi [\mnras] {10.1093/mnras/238.3.897},
  \href {https://ui.adsabs.harvard.edu/abs/1989MNRAS.238..897L} {238, 897}

\bibitem[\protect\citeauthoryear{{Liske} et~al.,}{{Liske}
  et~al.}{2008}]{Liske+08:2008MNRAS.386.1192L}
{Liske} J.,  et~al., 2008, \mn@doi [\mnras] {10.1111/j.1365-2966.2008.13090.x},
  \href {https://ui.adsabs.harvard.edu/abs/2008MNRAS.386.1192L} {386, 1192}

\bibitem[\protect\citeauthoryear{{Lyke} et~al.,}{{Lyke}
  et~al.}{2020}]{LykeSDSS16q:2020ApJS..250....8L}
{Lyke} B.~W.,  et~al., 2020, \mn@doi [\apjs] {10.3847/1538-4365/aba623}, \href
  {https://ui.adsabs.harvard.edu/abs/2020ApJS..250....8L} {250, 8}

\bibitem[\protect\citeauthoryear{{Marconi} et~al.,}{{Marconi}
  et~al.}{2022}]{ANDES_MArconi2022}
{Marconi} A.,  et~al., 2022, in {Evans} C.~J.,  {Bryant} J.~J.,   {Motohara}
  K.,  eds,  Society of Photo-Optical Instrumentation Engineers (SPIE)
  Conference Series Vol. 12184, Ground-based and Airborne Instrumentation for
  Astronomy IX. p. 1218424, \mn@doi{10.1117/12.2628689}

\bibitem[\protect\citeauthoryear{{Martins}, {Martinelli}, {Calabrese}  \&
  {Ramos}}{{Martins} et~al.}{2016}]{Martins2016}
{Martins} C.~J.~A.~P.,  {Martinelli} M.,  {Calabrese} E.,   {Ramos}
  M.~P.~L.~P.,  2016, \mn@doi [\prd] {10.1103/PhysRevD.94.043001}, \href
  {https://ui.adsabs.harvard.edu/abs/2016PhRvD..94d3001M} {94, 043001}

\bibitem[\protect\citeauthoryear{{McMahon}, {Banerji}, {Gonzalez}, {Koposov},
  {Bejar}, {Lodieu}, {Rebolo}  \& {VHS Collaboration}}{{McMahon}
  et~al.}{2021}]{VHS:McMahon_2021}
{McMahon} R.~G.,  {Banerji} M.,  {Gonzalez} E.,  {Koposov} S.~E.,  {Bejar}
  V.~J.,  {Lodieu} N.,  {Rebolo} R.,   {VHS Collaboration} 2021, VizieR Online
  Data Catalog, \href {https://ui.adsabs.harvard.edu/abs/2021yCat.2367....0M}
  {p. II/367}

\bibitem[\protect\citeauthoryear{{Mor}, {Netzer}  \& {Elitzur}}{{Mor}
  et~al.}{2009}]{Mor2009}
{Mor} R.,  {Netzer} H.,   {Elitzur} M.,  2009, \mn@doi [\apj]
  {10.1088/0004-637X/705/1/298}, \href
  {https://ui.adsabs.harvard.edu/abs/2009ApJ...705..298M} {705, 298}

\bibitem[\protect\citeauthoryear{{Moresco} et~al.,}{{Moresco}
  et~al.}{2022}]{Moresco2022}
{Moresco} M.,  et~al., 2022, \mn@doi [Living Reviews in Relativity]
  {10.1007/s41114-022-00040-z}, \href
  {https://ui.adsabs.harvard.edu/abs/2022LRR....25....6M} {25, 6}

\bibitem[\protect\citeauthoryear{{Morrissey} et~al.,}{{Morrissey}
  et~al.}{2007}]{GALEX:Morrissey_2007}
{Morrissey} P.,  et~al., 2007, \mn@doi [\apjs] {10.1086/520512}, \href
  {https://ui.adsabs.harvard.edu/abs/2007ApJS..173..682M} {173, 682}

\bibitem[\protect\citeauthoryear{{Murphy}, {Kacprzak}, {Savorgnan}  \&
  {Carswell}}{{Murphy} et~al.}{2019}]{MMurphy:squad}
{Murphy} M.~T.,  {Kacprzak} G.~G.,  {Savorgnan} G. A.~D.,   {Carswell} R.~F.,
  2019, \mn@doi [\mnras] {10.1093/mnras/sty2834}, \href
  {https://ui.adsabs.harvard.edu/abs/2019MNRAS.482.3458M} {482, 3458}

\bibitem[\protect\citeauthoryear{{Onken}, {Wolf}, {Bian}, {Fan}, {Jeat Hon},
  {Raithel}, {Tisserand}  \& {Lai}}{{Onken}
  et~al.}{2021}]{Onken21:2021arXiv210512215O}
{Onken} C.~A.,  {Wolf} C.,  {Bian} F.,  {Fan} X.,  {Jeat Hon} W.,  {Raithel}
  D.,  {Tisserand} P.,   {Lai} S.,  2021, arXiv e-prints, \href
  {https://ui.adsabs.harvard.edu/abs/2021arXiv210512215O} {p. arXiv:2105.12215}

\bibitem[\protect\citeauthoryear{{Pepe} et~al.,}{{Pepe}
  et~al.}{2021}]{ESPRESSO_Pepe2021}
{Pepe} F.,  et~al., 2021, \mn@doi [\aap] {10.1051/0004-6361/202038306}, \href
  {https://ui.adsabs.harvard.edu/abs/2021A&A...645A..96P} {645, A96}

\bibitem[\protect\citeauthoryear{{Pier} \& {Krolik}}{{Pier} \&
  {Krolik}}{1993}]{Pier1993}
{Pier} E.~A.,  {Krolik} J.~H.,  1993, \mn@doi [\apj] {10.1086/173427}, \href
  {https://ui.adsabs.harvard.edu/abs/1993ApJ...418..673P} {418, 673}

\bibitem[\protect\citeauthoryear{{Planck Collaboration} et~al.,}{{Planck
  Collaboration} et~al.}{2020}]{Planck18}
{Planck Collaboration} et~al., 2020, \mn@doi [\aap]
  {10.1051/0004-6361/201833910}, \href
  {https://ui.adsabs.harvard.edu/abs/2020A&A...641A...6P} {641, A6}

\bibitem[\protect\citeauthoryear{{Prochaska}}{{Prochaska}}{2017}]{specDB}
{Prochaska} J.~X.,  2017, \mn@doi [Astronomy and Computing]
  {10.1016/j.ascom.2017.03.003}, \href
  {https://ui.adsabs.harvard.edu/abs/2017A&C....19...27P} {19, 27}

\bibitem[\protect\citeauthoryear{{Reis}, {Baron}  \& {Shahaf}}{{Reis}
  et~al.}{2019}]{ReisPRF:2019AJ....157...16R}
{Reis} I.,  {Baron} D.,   {Shahaf} S.,  2019, \mn@doi [\aj]
  {10.3847/1538-3881/aaf101}, \href
  {https://ui.adsabs.harvard.edu/abs/2019AJ....157...16R} {157, 16}

\bibitem[\protect\citeauthoryear{{Richards} et~al.,}{{Richards}
  et~al.}{2006}]{Richards2006}
{Richards} G.~T.,  et~al., 2006, \mn@doi [\apjs] {10.1086/506525}, \href
  {https://ui.adsabs.harvard.edu/abs/2006ApJS..166..470R} {166, 470}

\bibitem[\protect\citeauthoryear{{Sandage}}{{Sandage}}{1962}]{Sandage62}
{Sandage} A.,  1962, \mn@doi [\apj] {10.1086/147385}, \href
  {https://ui.adsabs.harvard.edu/abs/1962ApJ...136..319S} {136, 319}

\bibitem[\protect\citeauthoryear{{Schindler} et~al.,}{{Schindler}
  et~al.}{2019a}]{SDSSIncomplete_Schindler_PS:2019ApJ...871..258S}
{Schindler} J.-T.,  et~al., 2019a, \mn@doi [\apjs] {10.3847/1538-4365/ab20d0},
  \href {https://ui.adsabs.harvard.edu/abs/2019ApJS..243....5S} {243, 5}

\bibitem[\protect\citeauthoryear{{Schindler} et~al.,}{{Schindler}
  et~al.}{2019b}]{SDSSIncomplete_Schindler:2019ApJ...871..258S}
{Schindler} J.-T.,  et~al., 2019b, \mn@doi [\apj] {10.3847/1538-4357/aaf86c},
  \href {https://ui.adsabs.harvard.edu/abs/2019ApJ...871..258S} {871, 258}

\bibitem[\protect\citeauthoryear{{Sevilla-Noarbe} et~al.,}{{Sevilla-Noarbe}
  et~al.}{2021}]{DESY3Gold}
{Sevilla-Noarbe} I.,  et~al., 2021, \mn@doi [\apjs] {10.3847/1538-4365/abeb66},
  \href {https://ui.adsabs.harvard.edu/abs/2021ApJS..254...24S} {254, 24}

\bibitem[\protect\citeauthoryear{{Skrutskie} et~al.,}{{Skrutskie}
  et~al.}{2006}]{2MASS:2006AJ....131.1163S}
{Skrutskie} M.~F.,  et~al., 2006, \mn@doi [\aj] {10.1086/498708}, \href
  {https://ui.adsabs.harvard.edu/abs/2006AJ....131.1163S} {131, 1163}

\bibitem[\protect\citeauthoryear{{Sun} \& {Malkan}}{{Sun} \&
  {Malkan}}{1989}]{Sun1989}
{Sun} W.-H.,  {Malkan} M.~A.,  1989, \mn@doi [\apj] {10.1086/167986}, \href
  {https://ui.adsabs.harvard.edu/abs/1989ApJ...346...68S} {346, 68}

\bibitem[\protect\citeauthoryear{{Vanden Berk} et~al.,}{{Vanden Berk}
  et~al.}{2001}]{VandenBerk2001}
{Vanden Berk} D.~E.,  et~al., 2001, \mn@doi [\aj] {10.1086/321167}, \href
  {https://ui.adsabs.harvard.edu/abs/2001AJ....122..549V} {122, 549}

\bibitem[\protect\citeauthoryear{{V{\'e}ron-Cetty} \&
  {V{\'e}ron}}{{V{\'e}ron-Cetty} \&
  {V{\'e}ron}}{2010}]{Veron10:2010A&A...518A..10V}
{V{\'e}ron-Cetty} M.~P.,  {V{\'e}ron} P.,  2010, \mn@doi [\aap]
  {10.1051/0004-6361/201014188}, \href
  {https://ui.adsabs.harvard.edu/abs/2010A&A...518A..10V} {518, A10}

\bibitem[\protect\citeauthoryear{{Wolf} et~al.,}{{Wolf}
  et~al.}{2018}]{SkyMapper1:2018PASA...35...10W}
{Wolf} C.,  et~al., 2018, \mn@doi [\pasa] {10.1017/pasa.2018.5}, \href
  {https://ui.adsabs.harvard.edu/abs/2018PASA...35...10W} {35, e010}

\bibitem[\protect\citeauthoryear{{Wolf} et~al.,}{{Wolf}
  et~al.}{2020}]{Wolf20QSO:2020MNRAS.491.1970W}
{Wolf} C.,  et~al., 2020, \mn@doi [\mnras] {10.1093/mnras/stz2955}, \href
  {https://ui.adsabs.harvard.edu/abs/2020MNRAS.491.1970W} {491, 1970}

\bibitem[\protect\citeauthoryear{{Wright} et~al.,}{{Wright}
  et~al.}{2010}]{WISE:2010AJ....140.1868W}
{Wright} E.~L.,  et~al., 2010, \mn@doi [\aj] {10.1088/0004-6256/140/6/1868},
  \href {https://ui.adsabs.harvard.edu/abs/2010AJ....140.1868W} {140, 1868}

\bibitem[\protect\citeauthoryear{{Yang} et~al.,}{{Yang}
  et~al.}{2016}]{paper:yangQSO}
{Yang} J.,  et~al., 2016, \mn@doi [\apj] {10.3847/0004-637X/829/1/33}, \href
  {https://ui.adsabs.harvard.edu/abs/2016ApJ...829...33Y} {829, 33}

\makeatother
\end{thebibliography}

%\appendix

\onecolumn
\begin{center}
\begin{longtable}{rcccccccc}
\caption{QUBRICS candidates with a new reliable spectroscopic identification derived from Gaia low-resolution spectra. When available, the redshift obtained with follow-up spectroscopy is reported, together with the date and the used spectrograph.  The $i$ magnitude is in the AB photometric system.
{\it Note: this is only the beginning of the table, as an example. The complete table will be published online.}
% and targets identified with * show broad absorption lines (BAL QSOs).
}
\label{tab:NewSpec} \\
\toprule
   qid       & RA            & DEC     & $i_{\rm psf}$ & z$_{\rm Gaia}$ &  Class & z$_{\rm spec}$  & Obs. date & Instrument \\
             & J2000         & J2000         &      &     &            &       &           &  \\
\midrule

%\begin{tabular}{|r|l|l|l|l|l|l|l|l|}
%\hline
%  \multicolumn{1}{|c|}{qid} &
%  \multicolumn{1}{c|}{RA} &
%  \multicolumn{1}{c|}{DEC} &
%  \multicolumn{1}{c|}{i\_tab\_2} &
%  \multicolumn{1}{c|}{z\_SCR\_tab} &
%  \multicolumn{1}{c|}{Type} &
%  \multicolumn{1}{c|}{z\_spec\_f} &
%  \multicolumn{1}{c|}{Obs\_Date} &
%  \multicolumn{1}{c|}{Instrument} \\
%\hline
  888971 & 00:01:23.91 & -58:57:22.2 & 17.982 & 3.255 & QSO & - & - & -\\
  889687 & 00:02:09.76 & -60:59:07.2 & 17.064 & 2.664 & QSO & - & - & -\\
  7896285 & 00:03:43.91 & -18:54:25.0 & 18.510 & 2.603 & QSO & - & - & -\\
  1039775 & 00:04:35.62 & -25:07:07.6 & 17.248 & 1.898 & QSO & - & - & -\\
  33517238 & 00:04:49.35 & -36:29:39.6 & 18.973 & 2.864 & QSO & - & - & -\\
  1169052 & 00:04:57.90 & -52:45:30.4 & 18.297 & 3.143 & QSO & - & - & -\\
  7930596 & 00:04:59.18 & -16:46:42.1 & 18.509 & 2.562 & QSO & - & - & -\\
  863587 & 00:05:02.01 & -59:25:13.1 & 17.858 & 2.421 & QSO & - & - & -\\
  1172305 & 00:05:29.01 & -46:55:21.7 & 18.405 & 2.579 & QSO & - & - & -\\
  1163878 & 00:06:28.87 & -48:31:32.0 & 18.369 & 3.602 & QSO & - & - & -\\
  33521462 & 00:07:05.63 & -47:26:29.5 & 18.998 & 2.560 & QSO & - & - & -\\
  33581625 & 00:07:19.11 & -34:51:31.3 & 18.941 & 2.403 & QSO & - & - & -\\
  7837051 & 00:08:55.29 & -23:17:02.6 & 18.513 & 3.434 & QSO & - & - & -\\
  33564048 & 00:09:19.75 & -31:10:26.4 & 18.696 & 2.753 & QSO & - & - & -\\
  1185342 & 00:09:46.32 & -21:24:34.1 & 18.510 & 2.778 & QSO & - & - & -\\
  33542027 & 00:11:15.89 & -47:24:41.9 & 18.833 & 2.462 & QSO & - & - & -\\
  888901 & 00:11:18.15 & -69:08:20.9 & 17.873 & 2.953 & QSO & - & - & -\\
  819906 & 00:12:59.29 & -12:42:29.4 & 17.938 & 2.736 & QSO & - & - & -\\
  1191530 & 00:13:29.11 & -36:00:50.4 & 18.589 & 2.414 & QSO & - & - & -\\
  33626537 & 00:14:24.74 & -63:31:29.4 & 18.877 & 2.606 & QSO & - & - & -\\
  33488458 & 00:16:10.44 & -38:47:11.7 & 18.726 & 2.958 & QSO & - & - & -\\
  8404090 & 00:17:08.70 & +14:17:43.5 & 17.723 & 2.824 & QSO & - & - & -\\
  33520136 & 00:17:29.41 & -60:32:45.7 & 18.703 & 2.658 & QSO & - & - & -\\
  33586920 & 00:17:42.84 & -37:14:12.5 & 18.384 & 2.493 & QSO & - & - & -\\
  1352951 & 00:18:29.95 & +03:19:03.3 & 16.595 & 2.777 & QSO & 2.786 & 2022-08-26 & DOLORES\\
  \dots \\
\bottomrule
\end{longtable}

\end{center}
%\input{TAB/NewSpec.tex}

% Don't change these lines
\bsp	% typesetting comment
\label{lastpage}
\end{document}